\documentclass[showpacs,preprintnumbers,amssymb,prd]{revtex4}
\usepackage{graphicx}
\usepackage{dcolumn}
\usepackage{color}
\usepackage{bm}

\begin{document}

\title{\bf Quantum entanglement in a non-commutative system}

\author{S. Adhikari}
\altaffiliation{satyabrata@bose.res.in}
\affiliation{S. N. Bose National Centre for Basic Sciences,
Salt Lake, Kolkata 700 098, India}
\author{B. Chakraborty}
\altaffiliation{biswajit@bose.res.in}
\affiliation{S. N. Bose National Centre for Basic Sciences,
Salt Lake, Kolkata 700 098, India}
\author{A. S. Majumdar}
\altaffiliation{archan@bose.res.in}
\affiliation{S. N. Bose National Centre for Basic Sciences,
Salt Lake, Kolkata 700 098, India}
\author{S. Vaidya}
\altaffiliation{vaidya@cts.iisc.ernet.in}
\affiliation{Centre For High Energy Physics, Indian Institute of Science,
Bangalore 560012, India}

%\date{today}

\begin{abstract}
We explore the effect of two-dimensional position-space
non-commutativity on the bipartite entanglement of continuous variable
systems. We first extend the standard symplectic framework for
studying entanglement of Gaussian states of commutative systems to
the case of non-commutative systems residing in two dimensions.
Using the positive partial transpose criterion for separability of
bipartite states we derive a new condition on the separability of
a non-commutative system that is dependent on the noncommutative parameter
$\theta$. We then consider the specific example of a
bipartite Gaussian state and show the quantitative reduction of entanglement
originating from non-commutative dynamics. We show that such a reduction of
entanglement for a non-commutative system arising from the modification of the
variances of the phase space variables (uncertainty relations)
is clearly manifested between two particles that are separated by
small distances.
\end{abstract}

\pacs{11.10.Nx,03.65.Ud}

\maketitle

\section{Introduction}

It is a common expectation that spacetime may not retain its smooth
continuous structure at very short distances, and hence can have
important consequences for physics at very high energies, like quantum
gravity. Moyal deformation of ordinary spacetime is an example of a
specific algebraic realization of this expectation \cite{dfr}.
Clarification of the role of underlying spacetime symmetries
\cite{Chaichian:2004za,aschieri} is crucial for
formulating quantum dynamics on such spaces \cite{bmpv,replyto,scgv}.
However, it is not only for considerations of quantum gravity that the
study of non-commutative spacetime is relevant: non-commutative
coordinates also arise in low-energy situations as well, like in the
Quantum Hall Effect\cite{qhe}. In the presence of a magnetic field, the
guiding centre coordinates of the electron become non-commutative.
Understanding the formulation of quantum theory on such a space
is important to address basic conceptual issues
like the behavior of many-particle
systems, notion of identity of particles \cite{bmpv,replyto}, and
connection between spin and statistics.

While it is not entirely surprising that spacetime non-commutativity
may lead to non-locality at short distances, we wish to address
another novel aspect of non-commutative quantum mechanics here. How are
notions of entanglement and information affected ? Quantum entanglement
is the key ingredient for
the storage and distribution of quantum information among the
fundamental constituents of the quantum world \cite{peres1}. The first
nontrivial consequence of entanglement on quantum ontology was noticed
many years ago within the context of position-momentum continuous variable
systems\cite{EPR}. In recent times there has been a rapid development
of the theory of entanglement for such systems\cite{reviews}. Several
interesting information processing applications
such as quantum teleportation\cite{bennett1}, dense-coding\cite{denscod}
and cryptography\cite{crypto} have been extended to the domain of
continuous variable entangled states\cite{teleport,cvcrypt}.

Understanding the physical origin of entanglement in quantum systems
has been a key goal of information theory, which is expected to
lead to insights on a vast range of diverse phenomena such as phase
transitions in condensed matter systems\cite{condmat} and black hole
physics\cite{bhinf}. Such a goal is unlikely to be completely
realized unless the role of the symplectic structure of the underlying
phase space on the information capacity or entropy of quantum systems
is clearly elicited. It seems likely that any modification to this
structure, as brought about by position-space non-commutativity at a
fundamental scale could impact the correlations of quantum
systems. There indeed should exist a deep-rooted connection between
non-commutative dynamics and the phenomenon of entanglement of quantum
states. Our motivation for this work is to initiate a systematic
investigation on the role of position-space noncommutativity on
entanglement. In the present study we restrict ourself to the
entanglement of bipartite Gaussian states for which the standard
symplectic formalism\cite{simon,simon1,duan} is readily available and
well-established in the commutative limit.

The plan of this paper is as follows. In Section IIA we briefly review the
standard symplectic formalism for determining the separability of
bipartite Gaussian states in a single spatial dimension. In
Section IIB we first motivate the need for considering
dynamics on a two-dimensional plane relevant for
non-commutative systems and then show how the variances
are modified under the effect of
non-commutativity leading to $\theta$-dependent uncertainty
relations. In Section IIC we extend the usual commutative
one-dimensional (spatial) formalism for application of the
positive partial transpose\cite{peres,horodecki} separability criterion for
Gaussian states\cite{simon1} to the case of a bipartite system in
two (spatial) dimensions.  Using the symplectic
formalism we derive the modified ($\theta$-dependent) condition
for the physicality of states on a non-commutative plane which leads
to a corresponding modification of the entanglement condition.
We then consider various examples of Gaussian states.
In Section IIIA the entanglement of a commutative bipartite system
in one (spatial) dimension is computed. In Section IIIB
a single particle on the two-dimensional noncommutative plane is studied
showing that the underlying non-commutativity does \emph{not}
impact the {\it a priori} separability of the modes in this case.
In Section IIIC an example of a bipartite state
in two dimensions obeying the usual commutative dynamics is considered
in order to illustrate the validity of
the symplectic formalism in two spatial dimensions. We finally
investigate a bipartite Gaussian state on a
two-dimensional non-commutative plane in Section IIID. We show that
the $\theta$-dependendent entanglement criterion leads to a reduction of
the magnitude of entanglement. We conclude with a
summary of our results in Section IV.

\section{Effect of non-commutative dynamics on entanglement of
  Gaussian states: formalism}

\subsection{Separability condition for a bipartite system}

To begin with, we will first review the symplectic formalism for
observing entanglement of a bipartite Gaussian state\cite{simon,simon1,duan}
where the usual position space commutativity is maintained. This
analysis is in order not only to make the paper self-contained, but
will be useful as a ready reference when we extend it to incorporate
non-commutativity in a subsequent section.  Let us consider a composite wave
function $\Psi$ describing the bipartite state of two particles
(in momentum space) as
\begin{eqnarray}
\Psi(p_1, p_2) = N_1[\psi_{1}(p_1)\otimes\psi_{2}(p_2) +
  \psi_{2}(p_1)\otimes\psi_{1}(p_2)]
\label{s1}
\end{eqnarray}
where $N_1$ is a normalization constant.

We have to first construct the variance matrix for the composite wave
function $\Psi$. We can then apply the positive partial transpose (PPT)
separability criterion.  The PPT criterion is a necessary condition for
separability of quantum states of arbitrary dimensions\cite{peres}.
It is also a sufficient condition for bipartite $2 \otimes 2$ and
$2 \otimes 3$ systems. Here we use the generalization of the PPT
criterion for Gaussian states\cite{simon1,duan} based on the smallest
symplectic eigenvalue
of the partially transposed variance matrix. The
variance matrix elements are calculated from the variances
and covariances of the phase-space operators
$\hat{x}_{1},\hat{p}_{1},\hat{x}_{2},\hat{p}_{2}$ in the state
$\Psi(p)$. The ordering of the phase-space variables in the variance
matrix for the state $\Psi(p)$ is chosen to be
$\hat{\xi}^{T}=(\hat{x}_{1},\hat{p}_{1},\hat{x}_{2},\hat{p}_{2})$.  The
$4\times4$ matrix formed by the set of bilinears
$\hat{\xi}^{\mu}\hat{\xi}^{\nu}$ transforms under symplectic
transformation as
\begin{eqnarray}
\{\hat{\xi}^{\mu}\hat{\xi}^{\nu}\} \to  \{\hat{\xi}'^{\mu}\hat{\xi}'^{\nu}\}
=\{S^{\mu}_{\alpha}S^{\nu}_{\beta}\hat{\xi}^{\alpha}\hat{\xi}^{\beta}\}
= S\{\hat{\xi}^{\alpha}\hat{\xi}^{\beta}\}S^{T},~~~~S \in
Sp(4,R)
\end{eqnarray}
where the indices $\mu, \nu, \alpha, \beta$ range from $1$ to $4$.
A general element of the operator matrix can be written in terms of
the commutator and anti-commutator as
\begin{eqnarray}
(\hat{\xi}\hat{\xi}^T)^{\mu\nu} &=&  \hat{\xi}^{\mu}\hat{\xi}^{\nu}\nonumber\\
&=& \frac{1}{2}\{\hat{\xi}^{\mu},\hat{\xi}^{\nu}\} +
  \frac{1}{2}[\hat{\xi}^{\mu},
\hat{\xi}^{\nu}]
\end{eqnarray}
where the real variance matrix $V$ for the state $\rho$ is
defined by
\begin{eqnarray}
\langle \hat{\xi}\hat{\xi}^T \rangle &=& {\rm Tr}(\rho
\hat{\xi}\hat{\xi}^T)\nonumber\\
&=& V + i\Omega
\end{eqnarray}

where
\begin{eqnarray}
V = \{v^{\mu\nu}\equiv\frac{1}{2} Tr (\rho\{\hat{\xi}^{\mu},
\hat{\xi}^{\nu}\})\}
\end{eqnarray}
can be identified as the variance matrix and the matrix
\begin{eqnarray}
i\Omega= \{i\Omega^{\mu\nu} \equiv \frac{1}{2} [\hat{\xi}^{\mu},
\hat{\xi}^{\nu}]\}~~~~~~~(\because \textrm{Tr}\rho=1)
\end{eqnarray}
is the symplectic matrix.
For a single mode system the variance matrix is given by
\begin{eqnarray}
V = \left[\begin{matrix}{\langle \hat{x}^2 \rangle & \frac{1}{2}
      \langle\{\hat{x},\hat{p}\} \rangle\cr
\frac{1}{2} \langle\{\hat{x},\hat{p}\} \rangle &  \langle\hat{p}^2
      \rangle}\end{matrix}\right]
\end{eqnarray}
For any matrix to be physically realizable (respecting the uncertainty
condition) one requires that
\begin{eqnarray}
\det V = \langle\hat{x}^2\rangle\langle\hat{p}^2\rangle - [\frac{1}{2}
  \langle \{\hat{x},\hat{p}\} \rangle]^2 \ge \frac{1}{4}
\end{eqnarray}
(in units of $\hbar = 1$). This is the symplectic invariant form of the
Heisenberg uncertainty
relation $\Delta \hat{x} \Delta \hat{p} \ge 1/2$, where we have
assumed without loss of generality that $\langle\hat{x}\rangle = 0 =
\langle\hat{p}\rangle$ by suitable translation in phase space, so that
$\langle\hat{x}^2\rangle = (\Delta \hat{x})^2$ and
$\langle\hat{p}^2\rangle = (\Delta \hat{p})^2$. Otherwise, the entries
of the variance matrix will take the form $V = \{v^{\mu\nu} \equiv
\frac{1}{2}\langle (\hat{\xi}^{\mu}\hat{\xi}^{\nu} +
\hat{\xi}^{\nu}\hat{\xi}^{\mu})\rangle - \langle\hat{\xi}^{\mu}\rangle
\langle\hat{\xi}^{\nu}\rangle \}$, which are clearly invariant under
phase space translations.  The variance matrix $V$ can be diagonalized
by suitable symplectic transformations to the form $V =
diag\{\nu/2,\nu/2\}$. The physicality condition then takes the form
\begin{eqnarray}
\nu^2 \ge 1
\end{eqnarray}

For an $n$-mode system, let us consider the state $\hat{\rho}(s) =
\hat{U}(s)\hat{\rho} \hat{U}^{\dagger}(s)$, where $\hat{U}$ is a
unitary operator implementing the symplectic transformation $S \in
Sp(2n,R)$. It then follows that \cite{simon}
\begin{eqnarray}
V' + i\Omega = S V S^T + i\Omega
\end{eqnarray}
It is clear that if $V$ is a physically realizable state, so is
its symplectic transformation $V'$ for every $S \in
Sp(2n,R)$. Further, using Williamson's theorem\cite{william}, it is
possible to transform the variance matrix $V$ using symplectic
transformations to the canonical scaled diagonal form, up to the
ordering of $\nu_j$, given by
\begin{eqnarray}
V' = S V S^T = diag
(\nu_1/2,\nu_2/2,\cdots,\nu_n/2,\nu_1/2,\nu_2/2,\cdots,\nu_n/2)
\end{eqnarray}
This set of eigenvalues $\{\nu_j\}$ which are at least doubly
degenerate, forms the symplectic spectrum of $V$ and can be
obtained alternatively by computing the ordinary eigenvalues of
$|2i\Omega V|$. One can then treat the $n$-mode system simply as
$n$ copies of the one-mode system, whereby $V'$ and $V$ are bona
fide (respecting the uncertainty relation) variance matrices, if
and only if (iff)
\begin{eqnarray}
\nu_j^2 \ge 1,  \>\> j=1,2,....n
\label{phys0}
\end{eqnarray}
The matrix given by $V' + i\Omega$ has a spectrum of eigenvalues
(usual eigenvalues, not symplectic eigenvalues)
$(\nu_j \pm 1)/2$ for $j=1,2...n$. It follows that the physicality
condition (\ref{phys0}) amounts to demanding that $V' + i\Omega
\ge 0$.  The symplectic transformation
\begin{eqnarray}
V' + i\Omega = S(V + i\Omega) S^T
\label{semidef}
\end{eqnarray}
connecting the two matrices $(V +i\Omega)$ and $(V' + i\Omega)$ preserves the
condition of positive semi-definiteness, i.e.,
\begin{eqnarray}
V + i\Omega \ge 0
\label{phys01}
\end{eqnarray}

In order to use the PPT criteria (as applied for
continuous variable systems\cite{simon1,duan}) for a two-particle
state, the variance matrix $V$ is transformed into
$\tilde{V}$ by applying time-reversal symmetry on the second
particle (partial transposition).  Equivalently, this is obtained by
flipping the sign of the momentum of the second particle so that the
variance matrix $V$ undergoes the following transformation
\begin{eqnarray}
V \rightarrow \tilde{V} = \Lambda V \Lambda \label{pt}
\end{eqnarray}
where $\Lambda= diag(1,1,1,-1)$.  We then again find the eigenvalues
$\tilde{\nu_{i}}$ of the matrix $|2i\Omega\tilde{V}|$, which form
the symplectic spectrum of the variance matrix $\tilde{V}$, where
$\Omega\equiv\oplus~\omega,~~\omega\equiv \left(\begin{matrix}{0& 1
\cr -1 &0 }\end{matrix}\right)$.  For separability of the state, the
PPT criterion reduces to a simple inequality satisfied by the smallest
symplectic eigenvalue $\tilde{\nu}_{-}$ of the partially transposed
state\cite{reviews}
\begin{eqnarray}
\tilde{\nu}_{-}^2 \geq 1
\end{eqnarray}
Any two-mode Gaussian state is separable if and only if the above
inequality holds, otherwise it is entangled.

The magnitude of entanglement can be estimated through the logarithmic
negativity\cite{logneg} defined as
\begin{eqnarray}
E = \mathrm{max}[0,-\mathrm{log}_2(\tilde{\nu}_{-})]
\label{logneg1}
\end{eqnarray}

\subsection{Kinematics on a non-commutative plane}

The algebra satisfied by the operator-valued coordinates
$\hat{\bar{x}}^{\mu}$ in the simplest form of non-commutative
space is given by
\begin{eqnarray}
[\hat{\bar{x}}^{\mu},\hat{\bar{x}}^{\nu}]=i\theta^{\mu\nu},
\label{nccom0}
\end{eqnarray}
where $\theta^{\mu\nu}$ is usually taken to be the
non-transforming entries of the constant antisymmetric matrix
$\Theta=\{\theta^{\mu\nu}\}$. In the usual $3+1$ dimensional
spacetime, the introduction of invariant length scales through
$\Theta$, however creates problem regarding Lorentz symmetry,
which can be restored only in a twisted framework in a
Hopf-algebraic setting \cite{Chaichian:2004za}. However, as it can
be easily seen that for the special case of a $2+1$ dimensional
spacetime, with time $t$ taken to be a commuting variable,
\begin{eqnarray}
[\hat{\bar{x}}_{1},\hat{\bar{x}}_{2}]=i\theta,~~~~[\hat{t},\hat{\bar{x}}_{1}]
=[\hat{t},\hat{\bar{x}}_{2}]=0\label{c100}
\end{eqnarray}
The entire set can be written covariantly (under $SO(2)$) as
\begin{eqnarray}
[\hat{\bar{x}}_{i},\hat{\bar{x}}_{j}]=i\theta\epsilon_{ij}=i\theta_{ij},
~~~~[\hat{t},\hat{\bar{x}}_{i}]=0~~~~~~ for~~ i,j=1,2 \label{c200}
\end{eqnarray}
where $\theta_{ij}=\theta\epsilon_{ij}$, ($\epsilon_{ij}$ is the
Levi-Civita alternating tensor with $\epsilon_{12}=+1$)
representing the spatial parts of the matrix $\Theta$, transforms
as a scalar under $SO(2)$. The space part of this $2+1$ dimensional
spacetime will henceforth be referred as non-commutative plane,
for which $\theta$ is the single non-commutative parameter
\cite{footnote1}. 

Our basic commutation relation therefore is
\begin{equation}
[\hat{\overline{x}}_i, \hat{\overline{x}}_j] = i \theta_{ij} = i \theta \epsilon_{ij}.
\label{nccom}
\end{equation}
Instead of working with operators $\hat{\bar{x}}_{i}$, one can work with
c-number valued coordinates $\bar{x}_{i}$, provided one
replaces operator products with Moyal star multiplication, which
is defined for a pair of generic functions $f(\bar{x}_{i})$ and
$g(\bar{x}_{i})$ belonging to Schwarz class \cite{szabo}
\begin{eqnarray}
f(\bar{x})\star
g(\bar{x})=f(\bar{x})\exp(\frac{i}{2}\theta\epsilon_{ij}
  \overleftarrow{\bar{\partial_{i}}}\overrightarrow{\bar{\partial_{i}}}) g(\bar{x}),
\quad \bar{\partial_{i}}\equiv\frac{\partial}{\partial
\bar{x}_{i}}
\end{eqnarray}
This can be expressed alternatively as
\begin{eqnarray}
f(\bar{x})\star g(\bar{x})=m_{\theta}(f(\bar{x})\otimes
g(\bar{x}))=m_{0}(\hat{F}_{\theta}f(\bar{x})\otimes g(\bar{x}))
\end{eqnarray}
where
$\hat{F}_{\theta}=\exp(\frac{-i}{2}\theta\epsilon_{ij}\hat{p}_{i}\otimes
\hat{p}_{j})$ is the Drinfeld twist operator 
and $m_{0}$ represents
pointwise multiplication i.e. $m_{0}(f(\bar{x})\otimes
g(\bar{x}))=f(\bar{x})g(\bar{x})$. With this, the Moyal bracket
\begin{eqnarray}
[\bar{x}_{i},\bar{x}_{j}]_{\star}=
\bar{x}_{i}\star\bar{x}_{j}-\bar{x}_{j}\star\bar{x}_{i}=i\theta\epsilon_{ij}
\end{eqnarray}
becomes isomorphic to Eq.(\ref{nccom}). This generates the
non-commutative algebra ${\cal A}_\theta({\mathbb
  R}^2)$ and represents a deformation of the corresponding commutative
algebra ${\cal A}_0({\mathbb R}^2)$ where the functions are composed through
point-wise multiplication. Because of this commutation relation, making boxes in
  noncommutative space is tricky at best, so we will avoid this
  route. Instead, we work directly with wave packets, characterized by
  its widths and average momenta/locations. But before doing so, we
  must first understand how to do quantum mechanics on the
  non-commutative plane.

In usual quantum theory, the $i^{th}$ coordinate position is represented by an
operator $\hat{x}_i$: it acts on a wavefunction $\psi(x)$  in the position
representation by multiplying $\psi(x)$ by $x_i$:
\begin{equation}
\hat{x}_i \psi(x) = x_i \psi(x) .
\end{equation}
For the non-commutative plane, we need to distinguish between left -
and right-multiplication operators:
\begin{eqnarray}
\hat{\overline{x}}^L_i \psi(\bar{x}) &=& \overline{x}_i \star \psi(\bar{x}), \\
\hat{\overline{x}}^R_i \psi(\bar{x}) &=& \psi(\bar{x}) \star
\overline{x}_i .
\end{eqnarray}
For concreteness, we shall henceforth choose $\hat{\overline{x}}^L_i$ as
operators corresponding to coordinate observables $\overline{x}_i$.

Our basic {\it quantum mechanical} commutation relations are:
\begin{eqnarray}
[\hat{\overline{x}}^L_i, \hat{\overline{x}}^L_j] &=& i \theta
  \epsilon_{ij} = -[\hat{\overline{x}}^R_i,
  \hat{\overline{x}}^R_j], \nonumber \\
{[}\hat{\overline{x}}^L_i, \hat{p}_j] &=& i  \delta_{ij}, \nonumber \\
{[}\hat{p}_i, \hat{p}_j] &=& 0 = [\hat{\overline{x}}^L_i,
  \hat{\overline{x}}^R_j] \label{ncqm}.
\end{eqnarray}
It is easy to check that
\begin{equation}
\hat{p}_k = \frac{1}{\theta} \epsilon_{km} (\hat{\overline{x}}^L_m -
\hat{\overline{x}}^R_m) \label{ncmom}
\end{equation}
satisfies the above commutation relations. In fact, $\hat{p}_k \psi(\overline{x})
= -i  \overline{\partial}_k \psi(\overline{x})$. Note that this is the same as
the adjoint action
of momentum introduced at the level of operators in \cite{scgv}.

Notice that
\begin{equation}
\hat{x}_i \equiv \frac{\hat{\overline{x}}^L_i + \hat{\overline{x}}^R_i}{2} \label{xcomm}
\end{equation}
form a commutative quantum algebra: $[\hat{x}_i, \hat{x}_j]=0$. So
the quantum operators $\hat{x}_i$ can be simultaneously
diagonalized, and their simultaneous eigenvalues generate the
commutative algebra ${\cal A}_0({\mathbb R}^2)$.

Also, one can easily check that $[\hat{x}_i, \hat{p}_j] = i
  \delta_{ij}$, so $\hat{x}_i$ and $\hat{p}_j$ generate the usual
  Heisenberg-Weyl (HW) algebra.
\begin{eqnarray}
\hat{x}_i \psi(x) &=& x_i \psi(x) \quad (\rm no \,\, star\,!), \\
\hat{p}_i \psi(x) &=& \frac{1}{\theta} \epsilon_{im} (\hat{\overline{x}}^L_m -
\hat{\overline{x}}^R_m) \psi(x) = -i  \overline{\partial}_i \psi(x).
\end{eqnarray}

Let
$\overrightarrow{\overline{\xi}}=\{\overline{\xi}~^{\alpha}\}=(\bar{x}_{1},\bar{x}_{2},p_{1},p_{2})^{T}$
be the corresponding phase-space coordinates.
The complete NC Heisenberg algebra is obtained by augmenting Eq.(\ref
{nccom}) by the following additional commutators
\begin{eqnarray}
[\hat{\bar{x}}_{i},\hat{p}_{j}]= i\delta_{ij},~~~~~~~~[\hat{p}_{i},\hat{p}_{j}]=0
\end{eqnarray}
Let the corresponding
underlying commutative structure be described by the phase-space
coordinates
$\overrightarrow{\xi}=\{\xi^{\mu}\}=(x_{1},x_{2},p_{1},p_{2})^{T}$
which satisfies the usual (commutative) Heisenberg algebra, i.e.,
for which $\theta=0$ in Eq.(\ref {nccom}). These phase-space
coordinates $\xi^{\mu}$ and $\overline{\xi}^{\alpha}$ are related
by the following linear transformation
\begin{eqnarray}
\xi^{\mu}= M^{\mu}~~_{\alpha}\overline{\xi}^{\alpha} \label{lt}
\end{eqnarray}
where the matrix $M$ is given by
\begin{eqnarray}
M= \{M^{\mu}~~_{\alpha}\}= \left(\begin{matrix}{1 & 0 & 0 &
\frac{\theta}{2} \cr 0 & 1 & \frac{-\theta}{2} & 0 \cr 0 & 0 &1 &
0 \cr 0 & 0 & 0 & 1 }\end{matrix}\right) \label{twoancilla}
\end{eqnarray}
Clearly, their NC counter-parts $\overline{V}$ and
$\overline{\Omega}$ are related to V and $\Omega$ by
\begin{eqnarray}
V=M\overline{V}M^{T} ~~~~~~~\textrm{and}
~~~~~~\Omega=M\overline{\Omega}M^{T}
\label{lt1}
\end{eqnarray}

Let us now consider the following
wave packet in momentum space:
\begin{eqnarray}
\psi(\vec{p};\vec{a},\vec{p}_{0},\alpha) = \frac{1}{\sqrt{\pi\alpha}}
e^{(-\frac{1}{2\alpha}(\vec{p} -
  \frac{\vec{p}_{0}}{2})^{2}-i\vec{p}\cdot \vec{b})}
\label{wp}
\end{eqnarray}
where $b_{i}=a_{i}+\frac{\theta\epsilon_{ik}p_{0k}}{4}$. The reason
for this particular choice of the form of $\vec{b}$ will become clear
subsequently.  The representation of $\hat{\overline{x}}_i$ in the
momentum space is given by
\begin{eqnarray}
\hat{\bar{x}}_{i}\psi(\vec{p};\vec{a},\vec{p_{0}},\alpha)\equiv
(i\partial_{p_{i}} - \frac{\theta\epsilon_{ij}p_{j}}{2})
\psi(\vec{p};\vec{a},\vec{p}_{0},\alpha)
\end{eqnarray}
We would now like to calculate the spreads in $\Delta\hat{
\bar{x}}_{1} ,\Delta\hat{ \bar{x}}_{2},\Delta \hat{p}_{1},\Delta
\hat{p}_{2}$ in the state $|\psi\rangle$ given by Eq.. (\ref
{wp}). From general arguments one expects that
\begin{eqnarray}
(\Delta\hat{ \bar{x}}_{1})(\Delta\hat{
\bar{x}}_{2})\geq\frac{\theta}{2}
\end{eqnarray}
\begin{eqnarray}
(\Delta\hat{ \bar{x}}_{1})(\Delta \hat{p}_{1})\geq\frac{1}{2}
\end{eqnarray}
\begin{eqnarray}
(\Delta\hat{ \bar{x}}_{2})(\Delta \hat{p}_{2})\geq\frac{1}{2}
\end{eqnarray}
To derive the uncertainty relations, we proceed with the variance
matrix approach.
The elements of the variance matrix in the basis of the
phase-space variables
$\hat{\bar{\xi}}^{\mu}\equiv\{\hat{x}_{1},\hat{x}_{2},\hat{p}_{1},\hat{p}_{2}\}$
in the non-commutative plane is obtained as  \cite{footnote3}
\begin{eqnarray}
V_{NC}^{1}=
\left(\begin{matrix}{\frac{\theta^{2}\alpha}{8}+\frac{1}{2\alpha}
& 0 & 0 & \frac{-\theta\alpha}{4} \cr 0 &
\frac{\theta^{2}\alpha}{8}+\frac{1}{2\alpha} &
\frac{\theta\alpha}{4} & 0 \cr 0 & \frac{\theta\alpha}{4} &
\frac{\alpha}{2} & 0 \cr \frac{-\theta\alpha}{4} & 0 & 0 &
\frac{\alpha}{2}}\end{matrix}\right) \label{vm2}
\end{eqnarray}
where
\begin{eqnarray}
V_{NC}^{(1)} = \{\langle\frac{(\hat{\bar{\xi}}^{\mu}\hat{\bar{\xi}}^{\nu} +
  \hat{\bar{\xi}}^{\mu}\hat{\bar{\xi}}^{\nu})}{2}\rangle_{\psi} -
\langle\hat{\bar{\xi}}^{\mu}\rangle_{\psi}\langle\hat{\bar{\xi}}^{\nu}\rangle_{\psi}\}
 ~~~~ as~~
\langle\hat{\bar{\xi}}^{\mu}\rangle_{\psi}\neq0, ~~\forall~
\hat{\bar{\xi}}^{\mu} \label{vme2}
\end{eqnarray}
For example, it turns out that
$\langle\hat{\vec{\overline{x}}}\rangle_{\psi}=\vec{a}$,
$\langle\hat{\vec{p}}\rangle_{\psi}=\frac{\vec{p}_{0}}{2}$, indicating that
the particle represents by packet (\ref {wp}) is located at $\vec{a}$
and has a mean momentum $\frac{\vec{p}_{0}}{2}$. These simple forms
for the mean position and momentum explains the choice of the
particular form of $\vec{b}$ in Eq.(\ref{wp}).

From the above variance matrix, we have
\begin{eqnarray}
\Delta \hat{\bar{x}}_1 = \Delta\hat{\bar{x}}_2 =
\sqrt{\frac{\theta^{2}\alpha}{8}+\frac{1}{2\alpha}}, \quad
\Delta\hat{p}_1 = \Delta\hat{p}_2 = \sqrt{\frac{\alpha}{2}}
\end{eqnarray}
Therefore, the "space-space" uncertainty in the state
$\psi(\vec{p};\vec{a},\vec{p}_0,\alpha)$ is given by
\begin{eqnarray}
\Delta \hat{\bar{x}}_1 \Delta\hat{\bar{x}}_2 =
\frac{\theta^{2}\alpha}{8} + \frac{1}{2\alpha}
\label{spread1}
\end{eqnarray}
This uncertainty is minimized when
$\alpha=\alpha_{0}=\frac{2}{\theta}$, and
\begin{eqnarray}
(\Delta\hat{\bar{x}}_1 \Delta\hat{\bar{x}}_2)_{min} = \frac{\theta}{2}
\label{minunc0}
\end{eqnarray}
It follows that the minimum space-space uncertainty cannot be
arbitrarily reduced below the value given by Eq.(\ref{minunc0}). This
is different from the case of the standard commutative dynamics
($\theta = 0$) where by increasing $\alpha$ (the spread of the wave
packet in momentum space), one can measure the observables $\hat{x}_1,\hat{x}_2$
to arbitrary precision.  Similarly, the "phase-space" uncertainties in
the state $\psi(\vec{p};\vec{a},\vec{p}_0,\alpha)$ is given by
\begin{eqnarray}
\Delta\hat{\bar{x}}_1 \Delta\hat{p}_1 = \Delta\hat{\bar{x}}_2
\Delta\hat{p}_2 = \sqrt{\frac{\theta^{2}\alpha^{2}}{16} + \frac{1}{4}}
\label{spread2}
\end{eqnarray}
For $\alpha=\alpha_{0}$, the ``phase-space'' uncertainties reduce to
$\Delta\hat{\bar{x}}_1 \Delta\hat{p}_2 = \Delta\hat{\bar{x}}_2
\Delta\hat{p}_2 = \frac{1}{\sqrt{2}}$.  The above equations
(\ref{spread1}) and (\ref{spread2}) determine the space-space and
phase-space uncertainties under the effect of non-commutative
dynamics.  For instance, the variances are clearly enhanced in the
non-commutative case ($\theta \neq 0$).  This suggests that by
increasing the uncertainty, i.e., the fuzziness of the unit cells in
phase space, the effect of non-commutativity is to decrease the
available degrees of freedom (the total number of unit cells) if the
overall volume of the phase space is held fixed.  The number of
allowed states in a noncommutative system is essentially
$ \sim \frac{A}{\theta}$ for a planar system of large but
finite area $A$.  The fact that this number is reduced in presence of
non-commutativity, was utilized to address the issue of fractional
Quantum Hall Effect in Ref.\cite{fqhe2} in a fermionic system involving electrons,
where the entropy gets reduced
generically, which also displays non-extensive behaviour \cite{scholtz}. One therefore
expects a similar reduction in entropy in the bosonic case also, just on the
basis of enhanced phase-space uncertainity (\ref{spread2}), which in turn is expected to
play a vital role in the entanglement property as well.
The consequences on the entanglement of two particles
will be evident later when we present explicit computations of the
entanglement of bipartite systems in the presence of non-commutative
dynamics.

Note that the commutative $\theta\rightarrow 0$ limit of the
variance matrix $V_{NC}^{1}$ reduces to
$V_{C}^{1}=diag(\frac{1}{2\alpha},\frac{1}{2\alpha},\frac{\alpha}{2},\frac{\alpha}{2})$.
The same form of this commutative variance matrix $V_{C}^{1}$ can
also be obtained alternatively from $V_{NC}^{1}$ by making use of
the linear transformation (\ref{lt1}) to express it in terms of the
commutative coordinates. The various uncertainty relations
can now be read off from this as
\begin{eqnarray}
\Delta \hat{x}_{1}\Delta \hat{x}_{2}=\frac{1}{2\alpha},~~ \Delta
\hat{x}_{1}\Delta \hat{p}_{1}=\Delta \hat{x}_{2}\Delta
\hat{p}_{2}=\frac{1}{2} \label{uncer.}
\end{eqnarray}
Clearly, by taking the spread $\alpha$ of the momentum space
Gaussian wave packets  arbitrarily large
$\alpha\rightarrow\infty$, we can find $(\Delta \hat{x}_{1}\Delta
\hat{x}_{2})_{min}=0$. This disappearance of the $\theta-dependence$
from $V_{C}^{1}$, or the various uncertainty relations following
from it, is a reflection of the fact that one does not expect to
observe any non-commutative effect at the single particle level, if the
commuting variables are used. We shall see subsequently that
this feature does not persist in the presence of two or more
particles, as the symplectic eigenvalues of the effective
commutative variance matrix will start displaying $\theta$-dependence
(non-commutative effect), thus
inheriting the original non-commutative features. However, in the
limit of $\alpha\rightarrow\infty$, the space-space uncertainty relations involving
non-commutative coordinates diverges (\ref{spread1}): $(\Delta
\hat{\overline{x}}_{1}\Delta \hat{\overline{x}}_{2})\rightarrow\infty$, which is
minimized only at $\alpha_{0}=\frac{2}{\theta}$ as we have seen
earlier.

\subsection{Entanglement of two particles on a non-commutative plane}

For a commutative system the correlation properties of the
two-dimensional Gaussian state is completely determined by an
$8\times8$ real symmetric variance matrix $V$. To study the
correlation properties it is convenient to transform the variance
matrix $V$ to some standard form through symplectic transformations.
Let us choose the basis vector for the phase space variables
to be $\vec{\xi} = (x_1^{(1)},
p_1^{(1)}, x_2^{(1)}, p_2^{(1)}, x_1^{(2)}, p_1^{(2)}, x_2^{(2)},
p_2^{(2)})$, where the subscripts $1$ and $2$ refer to the $x$ and $y$
components, respectively, and the superscripts $1$ and $2$ (within
parenthesis) refer to the first and second particle, respectively.
Any variance matrix can be written in terms of $4\times4$ block
matrices $\alpha$,~$\beta$ and $\gamma$ as
\begin{eqnarray}
\textsl{V} = \left(\begin{matrix}{\alpha & \gamma\cr \gamma^{T} &
\beta }\end{matrix}\right)_{8\times8} \label{varmat}
\end{eqnarray}
where
\begin{eqnarray}
\alpha =  \left(\begin{matrix}{A_{11} & B_{11} & A_{12} & B_{12}
\cr B_{11} & E_{11} & B_{21} & E_{12} \cr A_{12} & B_{21} & A_{22}
& B_{22} \cr B_{12} & E_{12} & B_{22} & E_{22}
 }\end{matrix}\right), ~~~~ \beta=
\left(\begin{matrix}{A'_{11} & B'_{11} & A'_{12} & B'_{12} \cr
B'_{11} & E'_{11} & B'_{21} & E'_{12} \cr A'_{12} & B'_{21} &
A'_{22} & B'_{22} \cr B'_{12} & E'_{12} & B'_{22} & E'_{22}
 }\end{matrix}\right)
\label{mat1}
\end{eqnarray}
represent the single particle (co-)variances for the first and second
particle, respectively, and
\begin{eqnarray}
\gamma = \left(\begin{matrix}{C_{11} & D_{11} & C_{12} & D_{12}
\cr D_{11} & G_{11} & D_{21} & G_{12} \cr C_{12} & D_{21} & C_{22}
& D_{22} \cr D_{12} & G_{12} & D_{22} & G_{22}
 }\end{matrix}\right)
\label{mat2}
\end{eqnarray}
represents the correlations between the two particles.

We now extend the procedure presented in Ref.\cite{duan} to the case
of a (spatial) two-dimensional system to transform any general
variance matrix (\ref{varmat}) to a standard symplectic invariant
form.  (Refer to Appendix I for details of this procedure).
The standard form
of the variance matrix is given by
\begin{eqnarray}
\textsl{V} = \left(\begin{matrix}{g_{a} & 0 & 0 & 0& m_{a}& 0 &
q_{a}& 0 \cr 0 & g_{a} & 0 & 0 & 0 & m_{c} & 0 & q_{b} \cr 0 & 0 &
g_{b} & 0& q_{a} & 0 & m_{b} &0  \cr 0 & 0 & 0 & g_{b}& 0 & q_{b}
& 0 & m_{d} \cr m_{a} & 0 & q_{a} & 0 & g_{c} & 0 & 0 & 0 \cr 0 &
m_{c} & 0 & q_{b} & 0 & g_{c} & 0 & 0 \cr q_{a} & 0 & m_{b} & 0 &
0 & 0 & g_{d} & 0 \cr 0 & q_{b}& 0 & m_{d} & 0 & 0 & 0 & g_{d}
 }\end{matrix}\right)_{8\times8}
\label{twostandard1}
\end{eqnarray}
A Gaussian state described by the variance matrix $V$ is physical
if the smallest symplectic eigenvalue $\nu_{-}^{C}\geq1$ with the
constraint $V\geq0$\cite{simon1,duan,reviews}.

The stage is now set to formulate our prescription for studying
the entanglement of a non-commutative system. To discuss the
entanglement properties of two identical particles in
two-dimensional non-commutative space, we must first define such
states appropriately. We briefly recall here the arguments
provided in \cite{bmpv,replyto} that give the correct
construction.

In noncommutative spacetime, generators of symmetry
transformations act differently, because this action has to be
compatible with the star-product. The correct implementation of,
say, the rotation operator $\hat{J}$ requires that $\hat{J}$ act on a
two-particle state not via the usual product $\hat{\Delta}_0(\hat{J}) = {\bf
\hat{1}} \otimes \hat{J} + \hat{J} \otimes {\bf \hat{1}}$ but through the deformed
co-product defined as, $\hat{\Delta}_\theta (\hat{J}) = \hat{F}_\theta^{-1} \hat{\Delta}_0
(\hat{J}) \hat{F}_\theta$, where $\hat{F}_\theta^{-1} =
\exp[(\frac{i\theta_{ij}}{2})(\hat{p}_{i}\otimes \hat{p}_{j})]$  is the
inverse of the twist operator.

In usual quantum mechanics of identical particles, all observables commute with the statistics (or flip) operator $\hat{\tau}_0$ that exchanges the two particles: $\hat{\tau}_0$ is {\it superselected}. The twisted two-particle rotation operator $\hat{\Delta}_\theta (\hat{J})$ does not commute with $\hat{\tau}_0$, but does so with the twisted flip operator $\hat{\tau}_\theta = \hat{F}^{-1}_\theta \hat{\tau}_0 \hat{F}_\theta = \hat{F}^{-2}_\theta \hat{\tau}_0$ (An easy way to verify the second equality is to look at the action of $\hat{\tau}_\theta$ on, say, plane waves).

To ensure that statistics remains superselected, we therefore require that the two-particle wave function be
\begin{equation}
\Psi= N(\psi_{1}\otimes\psi_{2} + \hat{F}_{\theta}^{-2} (\psi_{2}\otimes\psi_{1})) \label{sai}
\end{equation}
where $N$ is the normalization constant. Such a wavefunction transforms correctly under the action of spacetime symmetries (roughly speaking, the wavefunction
remains twist symmetric in all reference frames).

It has been shown recently that these twisted particles can have
some non-trivial consequences like violation of Pauli principle
\cite{pp} and energy shift in the system of degenerate electron
gas \cite{elgas}. Though $\theta_{ij} = \theta \epsilon_{ij}$
turns out to be an $SO(2)$ scalar as mentioned earlier and the
technique of twist symmetrization is motivated by its
generalizability to higher dimensions. Indeed, general arguments
coming from diffeomorphism invariance \cite{bgk} and quantum
integrability \cite{sv} support this.

How does one generalize the phase space operators to the two-particle sector? In standard quantum mechanics, the position operator in the two-particle sector is $\hat{\Delta}_0 (\hat{x}) = \hat{x} \otimes {\bf \hat{1}} + {\bf \hat{1}} \otimes \hat{x}$, of which we intuitively identify the first piece $\hat{x} \otimes {\bf \hat{1}}$ as the "position of the first particle". This identification is strengthened by the observation that on applying the flip operator $\hat{\tau}_0$ to this, we get ${\bf \hat{1}} \otimes \hat{x}$, which is the "position of the second particle". In the noncommutative case, the phase space operators are the exact analogues of these: they are given by $\hat{F}_{\theta}^{-1}({\bf \hat{1}} \otimes \hat{\overline{\xi}}^{\mu})\hat{F}_{\theta}$ and
$\hat{F}_{\theta}^{-1}(\hat{\overline{\xi}}^{\mu} \otimes {\bf \hat{1}})\hat{F}_{\theta}$, as one
can be obtained from the other by applying $\hat{\tau}_{\theta}$.

For notational convenience, one can formally divide the $(8\times 8)$
matrix into four blocks, i.e.,
\begin{eqnarray}
{\overline{V}} = \left(\begin{matrix}{(\overline{V}_{1}\otimes
I)_{4\times4} & (mixed~~ terms)_{4\times4} \cr (mixed~~
terms)_{4\times4} & (I\otimes \overline{V}_{1})_{4\times4}
}\end{matrix}\right)_{8\times8} \label{twononcom}
\end{eqnarray}
The left-hand upper corner block and the right hand lower corner
block contains the elements of $\overline{V}_{1}\otimes I$ and
$I\otimes \overline{V}_{1}$, respectively. The other two symmetric
blocks contains the mixed terms. The above matrix ${V}$ is
obtained from the symmetric part of the matrix $\langle
\hat{F}_{\theta}^{-1}{\hat{\xi}\hat{\xi}^{T}}\hat{F}_{\theta}\rangle$, where
$\hat{\xi}^{T}=(\hat{\bar{x}}_{1}\otimes I,\hat{\bar{x}}_{2}\otimes I,\hat{p}_{1}\otimes
I,\hat{p}_{2}\otimes I,I\otimes \hat{\bar{x}}_{1},I\otimes
\hat{\bar{x}}_{2},I\otimes \hat{p}_{1},I\otimes \hat{p}_{2})$.

The important condition of positive semi-definiteness that is
preserved under symplectic transformations for commutative systems
(\ref{semidef}), however does not generally hold true for
non-commutative systems. This is because $\overline{\Omega}$ is
preserved only by the matrix $\overline{S} = M_2^{-1}SM_2$, i.e.,
\begin{eqnarray}
\overline{\Omega} = \overline{S}~\overline{\Omega}~\overline{S}^T
\end{eqnarray}
where $M_2$ is given by
\begin{eqnarray}
\left(\begin{matrix}{M\otimes I& 0 \cr 0 & I\otimes
M}\end{matrix}\right)
\label{twoanc}
\end{eqnarray}
with $M$ is defined in Eq.(\ref{twoancilla}) \cite{footnote4}.  Thus,
the strategy employed here will be to transform the variance matrix
$\overline{V}$, to the effective commutative variance matrix $V$ using
Eq.(\ref{lt1}), which also corresponds to the (co-)variances in terms
of observables, as discussed earlier in Section IIB.  The rest of the
analysis parallels the one used for commutative systems, discussed in
Section IIA. As a result of the transformation the non-commutative parameter
$\theta$ enters into the effective commutative variance matrix
elements. Our
prescription will be to
first transform a general form of the noncommutative variance matrix to
an effective commutative variance matrix, and then use the standard form
for the latter to compute its symplectic
eigenvalues (see Eqs.(\ref{vm4})--(\ref{phys.y}) below).
In general, any non-commutative variance matrix of two particles when
transformed to the effective commutative plane will acquire
$\theta$-dependence.

It follows from Williamson's theorem\cite{william} that the effective
commutative variance matrix $V$ can be transformed by symplectic
transformations to the canonical scaled diagonal form where one can
separate out the $\theta$-dependence of the various terms. Thus, one
can write
\begin{eqnarray}
V' &=& S V S^T = S (V(\theta =0) + \delta V(\theta))S^T \nonumber\\
&=& diag (\nu_1(\theta=0) + \delta \nu_1(\theta),\cdots, \nu_4(\theta=0) +
\delta \nu_4(\theta), \nu_1(\theta=0) + \delta \nu_1(\theta),\cdots,
\nu_4(\theta=0) + \delta \nu_4(\theta))
\end{eqnarray}
In the commutative limit ($\theta=0$), the uncertainty relation
imposes the condition $\nu_j^2(\theta =0) \ge 1$ on the eigenvalues
(see, section IIA). It thus follows, that the effective variance
matrix $V$ is physical if and only if $(\nu_j (\theta=0) + \delta
\nu_j(\theta))^2 \ge 1 + (\delta \nu_j(\theta))^2 +2\nu_j
(\theta=0).\delta \nu_j(\theta)$ Now, consider the matrix $V' +
i\Omega$. It can be shown that this matrix has a spectrum of
eigenvalues given by $[\nu_j(\theta =0) + \delta \nu_j(\theta) \pm
1]/2$, for $j=1,2,3,4$.  Since the matrices $V' + i\Omega$ and $V+
i\Omega$ are related through a symplectic transformation, it follows
that the positivity condition is preserved, i.e.,
\begin{eqnarray}
V + i\Omega \ge 0
\end{eqnarray}
if $\delta \nu_j(\theta) > 0$ which turns out to be true in the example
we are considering here.

We therefore transform the non-commutative variance matrix into the
effective commutative variance matrix using the relations (\ref{lt1}).
We then proceed in the same way as for the commutative system.
Due to the $\theta$-dependence of the symplectic eigenvalues of the
effective commutative variance matrix, the condition analogous to
the physicality condition (\ref{phys0}) for the commutative case
is modified to a $\theta$-dependent relation here, given by
\begin{eqnarray}
(\nu_{-}^{NC})^{2} \ge (\nu_{-}^{NC}(\theta))^2_{min}
\label{modphys}
\end{eqnarray}
where $(\nu_{-}^{NC}(\theta))_{min} $ is the lowest symplectic eigenvalue
of the effective commutative variance matrix when the space-space
uncertainty $\Delta \hat{x}_1 \Delta \hat{x}_2 (\theta)$ is minimized.
Note that $(\nu_{-}^{NC})_{min} = 1$
when $\theta=0$ (since $\Delta \hat{x}_1 \Delta \hat{x}_2 = 0$ in this case),
i.e., for the commutative case,
one recovers back the usual physicality condition for commutative
systems.  As a result the similar condition on the smallest symplectic
eigenvalue of the partially transposed variance matrix (PPT criterion)
also gets modified, showing that the entanglement criterion for
noncommutative systems
\begin{eqnarray}
(\tilde{\nu}_-^{NC})^2 < (\nu_{-}^{NC}(\theta))^2_{min}
\label{entnoncom}
\end{eqnarray}
also acquires a $\theta$-dependence
We will derive the explicit form for $(\nu_{-}^{NC}(\theta))_{min} $ in the context
of the example that we will study in Section IIID.

\section{Computation of entanglement: some examples of Gaussian states}

In this section we will provide examples of specific Gaussian states
for which bipartite entanglement will be computed using the formalism
presented in the previous section. Though our ultimate goal is to
study the effect of non-commutativity on the entanglement of two
particles, we will proceed towards this direction step-by-step.
Hence, following the sequence of the previous section, we will proceed
by considering examples of entanglement in usual commutative dynamics
in one and two dimensions, before launching into the non-commutative
computations. This is in order to illustrate systematically the
formalism provided in the previous section.

\subsection{Two particles in one dimension obeying usual commutative dynamics}

Let us consider two particles in one dimension described by the two
normalized wave functions $\psi_{1}$ and $\psi_{2}$ given
respectively, by
\begin{eqnarray}
\psi_{1}(p) &=& \frac{1}{(\pi\alpha)^{\frac{1}{4}}}
\exp(\frac{-(p-p_{0}/2)^{2}}{2\alpha}) , \\
\psi_{2}(p) &=& \frac{1}{(\pi\beta)^{\frac{1}{4}}}
\exp(\frac{-(p+p_{0}/2)^{2}}{2\beta})
\end{eqnarray}
where $\alpha$ and $\beta$ denote respectively the spread of the two
wave packets.  If we now construct a joint two-particle state given by
Eq.(\ref{s1}), with $N_1^2 = [2(1 + \frac{2\sqrt{\alpha\beta}
\exp(\frac{-p_{0}^{2}}{\alpha + \beta})}{\alpha + \beta})]^{-1}$ as
the normalization constant, one can observe that this $\Psi(p_1,p_2)$
two-particle state cannot be written as a product of two wave
functions of the single particle states unless the parameters $\alpha$
and $\beta$ are equal and $\vec{p_{0}}=0$. Therefore the state
$\Psi(p_1,p_2)$ is entangled when $\alpha\neq \beta$ or when $\alpha =
\beta$ and $p_{0}\neq0$.

Explicitly the variance matrix for two particles in the state
$\Psi(p_1,p_2)$ in Eq.(\ref{s1}) is obtained as
\begin{eqnarray}
V = \left(\begin{matrix}{v_{11} & v_{12} & v_{13} & v_{14}
\cr v_{12} & v_{22} & v_{23} & v_{24} \cr v_{13} & v_{23} & v_{33}
& v_{34} \cr v_{14} & v_{24} & v_{34} & v_{44}}\end{matrix}\right)
\equiv \left(\begin{matrix}{\alpha' & \gamma' \cr \gamma'^{T} & \beta'
}\end{matrix}\right) \label{vm1}
\end{eqnarray}
where $\alpha',\beta',\gamma'$ are $2\times2$ block matrices.  Using
the tensor product notation for $\hat{\xi}^{T}=(\hat{x}\otimes
I,\hat{p}\otimes I,I \otimes\hat{x},I \otimes\hat{p})$, one gets
\begin{eqnarray}
&&v_{11}= v_{33}=\langle \hat{x}^{2}\otimes I\rangle_{\Psi}=\langle I
\otimes \hat{x}^{2}\rangle_{\Psi}= N_1^{2}\left(\frac{\alpha +
\beta}{2\alpha\beta} +
\frac{4\sqrt{\alpha\beta}\exp(\frac{-p_{0}^{2}}{\alpha +
  \beta})}{(\alpha +
    \beta)^2}(1-\frac{p_{0}^{2}}{\alpha + \beta})\right){}\nonumber\\
&& v_{12} = v_{34} = \langle (\hat{x}~\hat{p} + \hat{p}~\hat{x})/2\otimes
I\rangle_{\Psi} = \langle
I\otimes(\hat{x}~\hat{p}+\hat{p}~\hat{x})/2\rangle_{\Psi} = 0{}\nonumber\\
&&v_{13} = \langle \hat{x}\otimes \hat{x}\rangle_{\Psi} = N_1^2 \left(\frac{4 p_{0}^{2}
  \alpha\beta \exp(\frac{-p_{0}^{2}}{\alpha +
    \beta})}{\sqrt{\alpha\beta}(\alpha+\beta)^{3}}\right){}\nonumber\\
&&v_{14} = v_{23} = \langle \hat{x}\otimes \hat{p}\rangle_{\Psi}=\langle \hat{p}\otimes
\hat{x}\rangle_{\Psi}=0{}\nonumber\\
&& v_{22} = v_{44} = \langle \hat{p}^{2}\otimes I\rangle_{\Psi} = \langle I\otimes
\hat{p}^{2}\rangle_{\Psi}=N_1^{2} \left(\frac{\alpha + \beta}{2} +
  \frac{p_{0}^{2}}{2} +
  \frac{4\sqrt{\alpha\beta}\exp(\frac{-p_{0}^{2}}{\alpha +
      \beta})}{\alpha + \beta}(\frac{\alpha\beta}{\alpha + \beta} +
  \frac{p_{0}^{2}(\beta -\alpha)^{2}}{4(\alpha +
    \beta)^{2}})\right){}\nonumber\\
&& v_{24} = \langle \hat{p}\otimes \hat{p}\rangle_{\Psi} =
N_1^{2}\left(\frac{-p_{0}^{2}}{2} + \frac{p_{0}^{2}\sqrt{\alpha\beta}(\beta
    - \alpha)^{2} \exp(\frac{-p_{0}^{2}}{\alpha + \beta})}{(\alpha +
    \beta)^{3}}\right)
\label{velement}
\end{eqnarray}
Needless to say, the above variance matrix (\ref{vm1}) corresponds to a physical
state, and it can be easily verified that
the eigenvalues of $V$ satisfy the physicality condition $\nu_j \ge 1$.

In order to study the entanglement of the state
described by the variance matrix $V$,
we now perform a partial transposition of $V$ defined by Eq.(\ref{pt}).
The symplectic eigenvalues $\tilde{\nu}$ of $\tilde{V}$ (the ordinary
eigenvalues of the matrix $|2i\Omega \tilde{V}|$) can be expressed in terms
of the symplectic invariants $\det\alpha'$,  $\det\beta'$ and $\det\gamma'$ of
$\tilde{V}$, and are given by
\begin{eqnarray}
&&\tilde{\nu} = \pm \sqrt{2(\Delta\tilde{V} \pm
    \sqrt{(\Delta\tilde{V})^{2} - 4\det(\tilde{V})})} =
\pm 2\sqrt{(v_{11}\mp v_{13})(v_{22}\pm v_{24})}
\label{sse2par+1d}
\end{eqnarray}
where $\Delta\tilde{V}=\det\alpha' + \det\beta'-2 \det\gamma'$, and the
matrix elements $v_{11},v_{13},v_{22},v_{24}$ are given in
Eq.(\ref{velement}).  The smallest symplectic eigenvalue of the
partial transposed variance matrix $\tilde{V}$ is found out to be
\begin{eqnarray}
\tilde{\nu}_{-} &=& 2\sqrt{(v_{11}-v_{13})(v_{22}+v_{24})}{}\nonumber\\
&=& (\frac{1 + \eta}{1 + \eta + 2\sqrt{\eta} \exp(\frac{-\zeta}{1 +
    \eta})})\sqrt{\frac{1+\eta}{2\eta} +
\frac{4\sqrt{\eta}(1+\eta-2\zeta)\exp(\frac{-\zeta}{1+\eta})}{(1+\eta)^3}}\times{}\nonumber\\
&&\sqrt{\frac{1+\eta}{2} + \frac{\sqrt{\eta}(2\zeta(1 - \eta)^{2} +
    4\eta(1 +\eta)) \exp(-\zeta/(1 + \eta))}{(1 + \eta)^{3}}}
\label{case2}
\end{eqnarray}
where $\eta=\frac{\alpha}{\beta}$ and
$\zeta=\frac{p_{0}^{2}}{\beta}$.

\begin{figure}[h!]
\begin{center}
\includegraphics[width=12cm]{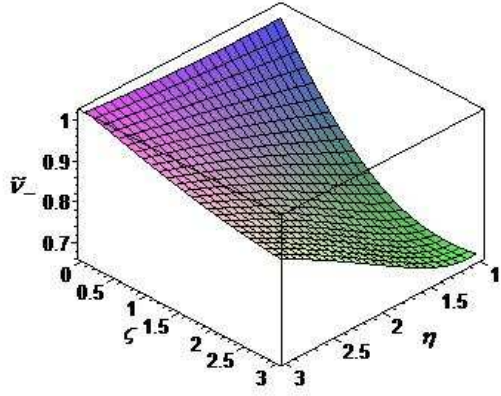}
\caption{(Coloronline) The smallest symplectic eigen value
$\tilde{\nu}_{-}$ is plotted versus the dimensionless parameters $\eta =
  \frac{\alpha}{\beta}$ and  $\zeta=\frac{p_{0}^{2}}{\beta}$.}
\end{center}
\label{f1}
\end{figure}

In Fig.1 we plot $\tilde{\nu}_{-}$ versus the two parameters
$\eta=\frac{\alpha}{\beta}$ and $\zeta=\frac{p_{0}^{2}}{\beta}$. From
the graph it is clear that the two-particle state is separable
$\tilde{\nu}_{-} = 1$ when $p_{0} = 0$ and $\alpha=\beta$. As
expected, the state is entangled for all other values of these
parameters.

\subsection{Single particle on a non-commutative plane}

Let us now inspect the wave function given by Eq.(\ref{wp}) for
any possible role of the underlying non-commutativity on its
separability property.  A single particle in two-dimensions can
also be viewed (if the issue of statistics is disregarded), as two
particles (or modes) in one-dimension, so that the wave function
(\ref{wp}) can be regarded as the direct product of two particles
on a line. On the basis of this argument, we can study the
separability or inseparability of the two-particle state described
by the effective commutative variance matrix $V_C^1$ in the para
above Eq.(\ref{uncer.}) in Section IIB.  Here our task is to find
the smallest symplectic eigenvalue $\tilde{\nu}_{-}$ of the
partially transposed state described by the transposed variance
matrix $(\tilde{V}_{C}^{1})$ obtained again from the variance
matrix $V_{C}^{1}$ using the transformation (\ref{pt}). The
smallest symplectic eigenvalue is found out to be $1$, i.e.,
$\tilde{\nu}_{-}=1$, which does not depend on the non-commutative
parameter $\theta$. Therefore, using the PPT criterion we can
conclude that the two-mode state described by the variance matrix
(\ref{vme2}) is separable for any non-commutative parameter
$\theta$, since as expected, non-commutativity does not play any
role at the single-particle level.

\subsection{Two particles moving on a two-dimensional
 commutative plane}

Let us consider two particles moving on a two dimensional plane
described by the wave functions
\begin{eqnarray}
\psi_{1}(\vec{p}_1) &=& \frac{1}{(\pi\alpha)^{\frac{1}{2}}}
\exp\left(-\frac{\vec{p}_1^{2}}{2\alpha}-i\vec{p}_1 \cdot
\vec{a}\right), \\
\psi_{2}(\vec{p}_2) &=& \frac{1}{(\pi\alpha)^{\frac{1}{2}}}
\exp\left(-\frac{\vec{p}_2^{2}}{2\alpha} + i\vec{p}_2 \cdot \vec{a}\right)
\end{eqnarray}
The composite wave function for the two particles is constructed to be
\begin{eqnarray}
\Psi(\vec{p_{1}},\vec{p_{2}}) =
N_2[\psi_{1}(\vec{p_1})\otimes\psi_{2}(\vec{p_2}) +
  \psi_{2}(\vec{p_1})\otimes\psi_{1}(\vec{p_2})]
\label{comp.1}
\end{eqnarray}
where $N_2^2=\frac{1}{2(1 + \exp(-2\alpha|\vec{a}|^{2}))}$ is the
normalization constant.

The variance matrix can be written in the covariant form as
\begin{eqnarray}
V^{(2)}= \left(\begin{matrix}{A_{ij} & B_{ij}& C_{ij} & D_{ij}\cr
B_{ji} & E_{ij} & D_{ji}& G_{ij}\cr C_{ij} & D_{ij} & A_{ij} &
B_{ij} \cr D_{ji} & G_{ij} & B_{ji} & E_{ij} }
\end{matrix}\right)_{8\times8}
\label{vm2dcom.}
\end{eqnarray}
where $A_{ij}$,$B_{ij}$,$C_{ij}$,$D_{ij}$,$E_{ij}$ and $G_{ij}$
are $2\times2$ block matrices given by
\begin{eqnarray}
A_{ij} &=& \langle \hat{x}_{i}~ \hat{x}_{j}\otimes I\rangle_{\Psi}-\langle
(\hat{x}_{i}\otimes I)\rangle_{\Psi}\langle (\hat{x}_{j}\otimes
I)\rangle_{\Psi}{}\nonumber\\
&=&\langle I\otimes \hat{x}_{i}~\hat{x}_{j} \rangle_{\Psi}-\langle (I\otimes
\hat{x}_{i})\rangle_{\Psi}\langle (I\otimes
\hat{x}_{j})\rangle_{\Psi}{}\nonumber\\
&=& N_2^{2}[2a_{i}a_{j} + \frac{\delta_{ij}}{\alpha} +
  \frac{\delta_{ij}}{\alpha} \exp(-2\alpha|\vec{a}|^2)], \\
B_{ij} &=& \langle (\frac{(\hat{x}_{i}\hat{p}_{j} +
  \hat{p}_{j}\hat{x}_{i})}{2} \otimes I)\rangle_{\Psi} - \langle
(\hat{x}_{i} \otimes I)\rangle_{\Psi}\langle \hat{p}_{j}\otimes
I\rangle_{\Psi}{}\nonumber\\
&=& \langle (I\otimes\frac{(\hat{x}_{i}\hat{p}_{j} +
  \hat{p}_{j}\hat{x}_{i})}{2}) \rangle_{\Psi}-\langle (I\otimes
\hat{x}_{i})\rangle_{\Psi}\langle I\otimes
\hat{p}_{j}\rangle_{\Psi}{}\nonumber\\
&=& 0, \\
C_{ij} &=& \langle \hat{x}_{i}\otimes \hat{x}_{j}\rangle_{\Psi}-\langle
(\hat{x}_{i}\otimes I)\rangle_{\Psi}\langle (I\otimes
\hat{x}_{j})\rangle_{\Psi}=-2N_2^{2}a_{i}a_{j}, \\
D_{ij} &=& \langle \hat{x}_{i}\otimes \hat{p}_{j})\rangle_{\Psi}-\langle
\hat{x}_{i}\otimes
I)\rangle_{\Psi}\langle I\otimes \hat{p}_{j}\rangle_{\Psi}=0, \\
E_{ij} &=& \langle \hat{p}_{i}\hat{p}_{j}\otimes I\rangle_{\Psi}-\langle
\hat{p}_{i}\otimes I\rangle_{\Psi}\langle \hat{p}_{j}\otimes
I\rangle_{\Psi}=\langle I\otimes \hat{p}_{i}\hat{p}_{j}\rangle_{\Psi}-\langle
I\otimes \hat{p}_{i})\rangle_{\Psi}\langle I\otimes
\hat{p}_{j}\rangle_{\Psi}{}\nonumber\\
&=& N_2^{2}[\alpha\delta_{ij} + (\alpha\delta_{ij} - 2a_{i}a_{j}
  \alpha^{2}) \exp(-2\alpha|\vec{a}|^{2})],\\
G_{ij} &=& \langle \hat{p}_{i}\otimes \hat{p}_{j})\rangle_{\Psi}-\langle
\hat{p}_{i}\otimes I \rangle_{\Psi}\langle I\otimes
\hat{p}_{j}\rangle_{\Psi}=
2N_2^{2}a_{i}a_{j}\alpha^{2} \exp(-2\alpha|\vec{a}|^{2})
\end{eqnarray}
To simplify the calculations and without any loss of generality we
can take $a_{2}=0$.
The $2\times2$ and $4\times4$ submatrices
whose determinants are invariant under symplectic transformation
are as follows:
\begin{eqnarray}
&&\alpha_{a}=\beta_{a}= \left(\begin{matrix}{A_{11} & B_{11}\cr
B_{11} & E_{11}}
\end{matrix}\right)_{2\times2},\alpha_{b}=\beta_{b}= \left(\begin{matrix}{A_{22} & B_{22}\cr B_{22} &
E_{22}}
\end{matrix}\right)_{2\times2},
\gamma_{a}= \left(\begin{matrix}{C_{11} & D_{11}\cr D_{11} &
G_{11}}
\end{matrix}\right)_{2\times2},{}\nonumber\\&&\gamma_{b}= \left(\begin{matrix}{C_{22} & D_{22}\cr D_{22} &
G_{22}}
\end{matrix}\right)_{2\times2},
\delta_{a}= \left(\begin{matrix}{\alpha_{a} & \gamma_{a}\cr
\gamma_{a}^{T} & \beta_{a}}
\end{matrix}\right)_{4\times4},\delta_{b}= \left (\begin{matrix}{\alpha_{b} &
\gamma_{b}\cr \gamma_{b}^{T} & \beta_{b}}
\end{matrix}\right)_{4\times4}
\label{twoancilla2}
\end{eqnarray}
The matrix elements $g_{a},g_{b},g_{c},g_{d},m_{a},m_{b},m_{c},m_{d}$
of the standard form of the matrix (\ref{twostandard1}) are determined
by the symplectic invariants
$\det(\alpha_{a})=\det(\beta_{a})=g_{a}^{2}=g_{c}^{2}$,
$\det(\alpha_{b})=\det(\beta_{b})=g_{b}^{2}=g_{d}^{2}$,
$\det(\gamma_{a})=m_{a}m_{c}$, $\det(\gamma_{b})=m_{b}m_{d}$ and
$\det(V)= \det(\delta_{a}).\det(\delta_{b})$.

Our task is to calculate the symplectic eigenvalues of the
matrix $V$. The eigenvalues of the matrix
$|2i\Omega V|$ are invariant under the action
of symplectic transformation on the matrix $V$. Therefore, the
symplectic eigenvalues $\nu_{i}$'s can be expressed in terms of
symplectic invariant quantities det$\alpha_{a}$, det$\alpha_{b}$,
det$\beta_{a}$, det$\beta_{b}$, det$\gamma_{a}$, det$\gamma_{b}$,
 det$\delta_{a}$, det$\delta_{b}$ and are given by
\begin{eqnarray}
\nu_{x} &=& \pm\sqrt{2\Delta_{x}(V)\pm 2 \sqrt{(\Delta_{x}(V))^{2} - 4
    \det\delta_{a}}} , \\
\nu_{y} &=& \pm\sqrt{2\Delta_{y}(V)\pm 2 \sqrt{(\Delta_{y}(V))^{2} - 4
    \det\delta_{b}}}
\end{eqnarray}
where $\Delta_{x}(V)=\det\alpha_{a}+\det\beta_{a}+2\det\gamma_{a}$ and
$\Delta_{y}(V)=\det\alpha_{b}+\det\beta_{b}+2\det\gamma_{b}$.  The
smallest symplectic eigenvalue for the commutative system is given by
\begin{equation}
\nu_{-}^{C} = min(\nu_{x_{-}}^{C},\nu_{y_{-}}^{C})
\label{sse1}
\end{equation}
where
\begin{eqnarray}
\nu_{x_{-}}^{C} &=& \sqrt{2\Delta_{x}(V)- 2 \sqrt{(\Delta_{x}(V))^{2}
    - 4 \det\delta_{a}}} \label{numinusx}, \\
\nu_{y_{-}}^{C} &=& \sqrt{2\Delta_{y}(V)- 2 \sqrt{(\Delta_{y}(V))^{2}
    - 4\det\delta_{b}}}
\end{eqnarray}

After taking partial transpose of the variance matrix $V^{(2)}$, The
smallest symplectic eigenvalue is given by
\begin{eqnarray}
\tilde{\nu}_{-}^{C}=min(\tilde{\nu}_{x_{-}}^{C},\tilde{\nu}_{y_{-}}^{C})
\label{comm.sse}
\end{eqnarray}
where
\begin{eqnarray}
\tilde{\nu}_{x_{-}}^{C} &=& 2\sqrt{(A_{11} + C_{11})(E_{11} -
  G_{11})}{}\nonumber\\
&=& \frac{1}{(1+ \exp(-2\alpha a_{1}^{2}))}\sqrt{(1 + \exp(-2\alpha
  a_{1}^{2}))(1+(1 - 4\alpha a_{1}^{2})exp(-2\alpha a_{1}^{2}))}
\label{comm.symp.x}, \\
\tilde{\nu}_{y_{-}}^{C} &=& 2\sqrt{A_{22}E_{22}}= 1
\label{comm.symp.y}
\end{eqnarray}

\begin{figure}[h!]
\begin{center}
\includegraphics[width=12cm]{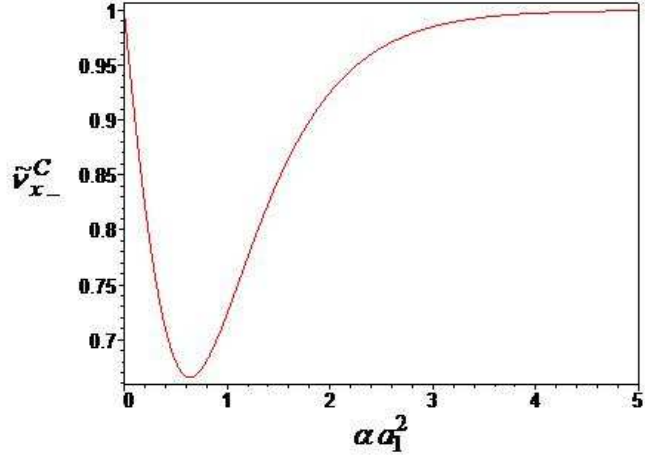}
\caption{(Coloronline) The symplectic eigenvalue
$\tilde{\nu}_{x_{-}}^{C}$ is plotted versus $\alpha a_{1}^{2}$.}
\end{center}
\label{f2}
\end{figure}

It is clear from Fig.2 that the smallest symplectic eigenvalue
$\tilde{\nu}_{-}^{C}$ is less than $1$ except at $a_{1}=0$ and
$a_{1}\rightarrow \infty$. For these two cases the smallest symplectic
eigenvalue takes the value $1$ and hence the state $\Psi(\vec{p})$
given in Eq. (\ref{comp.1}) is separable. For all other cases the
state is entangled, as is evident also from the structure of the wave
function in Eq.(\ref{comp.1}).

\subsection{Two particles on a non-commutative plane}

Let us finally consider the two particle state $\Psi$ relevant to
study the non-commutative dynamics on a two-dimensional plane:
\begin{eqnarray}
\Psi = N(\psi_1\otimes \psi_2 + \hat{F}_{\theta}^{-2}(\psi_2 \otimes \psi_1))
\label{realstate}
\end{eqnarray}
where $\hat{F}_{\theta}^{-1} = \exp\left((\frac{i\theta_{ij}}{2})(\hat{p}_i\otimes
\hat{p}_j)\right)$ is the (inverse) twist operator, as defined earlier in
Section II, and $N$ is a normalization constant.  $\psi_1$ and
$\psi_2$ are given by
\begin{eqnarray}
\psi_{1}=\psi_{1}(\vec{p};\vec{a},\vec{p}_0,\alpha)=\frac{1}{\sqrt{\pi\alpha}}
\exp((\frac{-1}{2\alpha})(\vec{p}-\frac{\vec{p}_0}{2})^{2}-i\vec{p}
\cdot \vec{a} -\frac{i\theta_{ij}p_{i}p_{0j}}{4})
\end{eqnarray}
and $\psi_{2} = \psi_{1}(\vec{p};-\vec{a},-\vec{p}_0,\alpha)$.

The variance matrix can be obtained after a long but straightforward
computation (see Appendix II and III) in the covariant form and is
given by
\begin{eqnarray}
\bar{V}= \left(\begin{matrix}{\bar{A}_{ij} & \bar{B}_{ij}&
\bar{C}_{ij} & \bar{D}_{ij}\cr \bar{B}_{ji} & \bar{E}_{ij} &
\bar{D}_{ji}& \bar{G}_{ij}\cr \bar{C}_{ij} & \bar{D}_{ij} &
\bar{A}_{ij} & \bar{B}_{ij} \cr \bar{D}_{ji} & \bar{G}_{ij} &
\bar{B}_{ji} & \bar{E}_{ij} }\end{matrix}\right)_{8\times8}
\label{vm4}
\end{eqnarray}
in the basis $\vec{\xi} = (\bar{x}_1^{(1)},\bar{x}_2^{(1)},p_1^{(1)},p_2^{(1)},\bar{x}_1^{(2)},\bar{x}_2^{(2)},p_1^{(2)},p_2^{(2)})$, where
$\bar{A}_{ij}$, $\bar{B}_{ij}$, $\bar{C}_{ij}$, $\bar{D}_{ij}$, $\bar{E}_{ij}$
and $\bar{G}_{ij}$ are $2\times2$ block matrices whose explicit forms
are given in Appendix II.

In order to investigate the presence of entanglement in the
two-particle state $\Psi$, described by the variance matrix
(\ref{vm4}), we have to again use the transformation (\ref{lt})
which transforms the variance matrix for the state in the
non-commutative plane to the effective variance matrix (corresponding to
the variance of commutative observables) in the commutative plane.
Thereafter, we
can apply the partial transposition to
detect the entanglement in the state $\Psi$ using the PPT criterion.
Now, the transformed variance
matrix in the commutative plane is given by
\begin{eqnarray}
V= \left(\begin{matrix}{A_{ij} & B_{ij}& C_{ij} & D_{ij}\cr B_{ji}
& E_{ij} & D_{ji}& G_{ij}\cr C_{ij} & D_{ij} & A_{ij} & B_{ij} \cr
D_{ji} & G_{ij} & B_{ji} & E_{ij} }
\end{matrix}\right)_{8\times8}
\label{vm5}
\end{eqnarray}
where $A_{ij}$, $B_{ij}$, $C_{ij}$, $D_{ij}$, $E_{ij}$ and $G_{ij}$
are $2\times2$ block matrices related to the earlier (barred)
matrices as follows:
\begin{eqnarray}
A_{ij}=\bar{A}_{ij}+\frac{\theta}{2}(\epsilon_{jk}\bar{B}_{ik}+\epsilon_{ik}\bar{B}_{jk})+
\frac{\theta^{2}}{4}\epsilon_{ik}\epsilon_{jl}\bar{E}_{kl}
\label{aij}
\end{eqnarray}
\begin{eqnarray}
B_{ij}=\bar{B}_{ij}+\frac{\theta}{2}\epsilon_{ik}\bar{E}_{kj}
\end{eqnarray}
\begin{eqnarray}
C_{ij}=\bar{C}_{ij}+\frac{\theta}{2}(\epsilon_{jk}\bar{D}_{ik}+
\epsilon_{ik}\bar{D}_{kj})+\frac{\theta^{2}}{4}\epsilon_{ik}\epsilon_{jl}\bar{G}_{kl}
\end{eqnarray}
\begin{eqnarray}
D_{ij}=\bar{D}_{ij}+\frac{\theta}{2}\epsilon_{ik}\bar{G}_{kj}
\end{eqnarray}
\begin{eqnarray}
G_{ij}=\bar{G}_{ij}
\end{eqnarray}
\begin{eqnarray}
E_{ij}=\bar{E}_{ij}
\end{eqnarray}

We first determine the physicality condition for the variance
matrix (\ref{vm5}), and
then perform the partial transposition of it.
Finally, we compute its symplectic eigenvalues.
For a straightforward illustration of the effect of non-commutativity
on entanglement, let us consider the following case.
We assume that the two particles described by the
wave functions $\psi_{1}$ and $\psi_{2}$ are in a static position
with respect to
each other,
i.e., $\vec{p}_{0}=0$ and that one of the
components of $\vec{b}$ (defined  as $b_i = a_i + \theta_{ij}p_{0j} = a_i$)
is zero,
(say, $b_2 = 0$).
With the above assumptions, the symplectic eigenvalues
$\nu_{x_{-}}^{NC}$ and $\nu_{y_{-}}^{NC}$ are obtained to be
\begin{eqnarray}
\nu_{x_{-}}^{NC} &=& 2\sqrt{(A_{11}-C_{11})(E_{11}-G_{11})}=
2\sqrt{((\bar{A}_{11} -
  \bar{C}_{11})+\frac{\theta^{2}}{4}(\bar{E}_{22} -
  \bar{G}_{22}))(\bar{E}_{11}-\bar{G}_{11})} {}\nonumber\\
&=&2N^{2}\sqrt{1 + \frac{3\theta^{2}\alpha^{2}}{4} + 4b_{1}^{2}\alpha
  + \frac{4(3\alpha^{2}\theta^{2}+4) \exp(\frac{-8\alpha
b_{1}^{2}}{(\alpha^{2}\theta^{2}+4)})}{(\alpha^{2}\theta^{2} + 4)^{2}}}
\times {}\nonumber\\
&&\sqrt{1+ \frac{16 \exp(\frac{-8\alpha
b_{1}^{2}}{(\alpha^{2}\theta^{2}+4)})}{(\alpha^{2} \theta^{2} +
    4)^{3}}(\theta^{2} \alpha^{2}-16\alpha b_{1}^{2}+4)}
\label{phys.x}, \\
\nu_{y_{-}}^{NC} &=& 2\sqrt{(A_{22}-C_{22})(E_{22}-G_{22})} =
2\sqrt{(\bar{A}_{22}-\bar{C}_{22}  + \frac{\theta^{2}}{4}(\bar{E}_{11}
  - \bar{G}_{11}))(\bar{E}_{22}-\bar{G}_{22})}{}\nonumber\\
&=& 2N^{2}\sqrt{1+\frac{3\theta^{2}\alpha^{2}}{4} +
  \frac{4 \exp(\frac{-8\alpha
b_{1}^{2}}{(\alpha^{2}\theta^{2}+4)})}{(\alpha^{2}\theta^{2} +
    4)^{3}}(3\theta^{4}\alpha^{4} + 16\alpha^{2}\theta^{2} + 16-
  32\theta^{2} \alpha^{3}b_{1}^{2})}\times {}\nonumber\\
&&\sqrt{1+ \frac{16 \exp(\frac{-8\alpha
b_{1}^{2}}{(\alpha^{2}\theta^{2}+4)})}{(\alpha^{2}\theta^{2} +
    4)^{2}}}  \label{phys.y}
\end{eqnarray}

The variance matrix (\ref{vm5}) represents the physical state if
it satisfies the appropriate physicality condition for a non-commutative
system explained in  section IIC, {\it viz.}
\begin{eqnarray}
(\nu_-^{NC})^2 = min ((\nu_{x_-}^{NC})^2,(\nu_{y_-}^{NC})^2) \ge (\nu_{-}^{NC}(\theta))^2_{min}
\label{physcon001}
\end{eqnarray}
In order to extract the $\theta$-dependence in the physicality
condition, we have to first obtain the minimum uncertainty state for
which the function $\Delta \hat{x}_1 \Delta \hat{x}_2 (\theta)$ is minimized.
The above function   can be read off from the elements of the effective
commutative variance matrix (\ref{vm5}), and is given by $A_{11}A_{22}$ in terms of
parameters defined in Eq.(\ref{aij}). Note that $(\nu_{-}^{NC}(\theta))_{min}$
depends also on the state parameters $\alpha$ and $b_1$. For the present
computation we minimize $\Delta \hat{x}_1 \Delta \hat{x}_2 (\theta)$ with
respect to the spread $\alpha$ for various values of the particle separation
$a_1 = b_1$. Or in other words, choosing a particular state corresponding
to a specific value of $b_1$, we compute the value of $\alpha$ (say, $\alpha_{min}$)
for which the uncertainty $\Delta \hat{x}_1 \Delta \hat{x}_2 (\theta)$ is minimized.
Next, we find the minimum symplectic eigenvalue  $(\nu_{-}^{NC}(\theta))_{min} $
by substituting the values of $\alpha_{min}$ and $b_1$ in Eqs.(\ref{phys.x})
and (\ref{phys.y}). We then repeat this computation
for other values of $b_1$.  We find that the value of $\alpha_{min}$ minimizing
the position-position
uncertainty stays constant for a specific range of the parameter $b_1$, and then
again acquires a different contant value for another range of $b_1$.
In this way we obtain the physicality condition
for ranges of states parametrized by various values of particle separation $b_1$.
The values of $(\nu_{-}^{NC}(\theta))_{min} $ corresponding to three
different ranges of values for the parameter $b_1$ are plotted in Figs.3, 4
and 5, with the respective minimum uncertainty conditions, i.e.,
$\alpha_{min} \simeq 1.4/\theta$,  $\alpha_{min} \simeq 1.6/\theta$ and
$\alpha_{min} \simeq 1.8/\theta$ displayed respectively on the figures.

\begin{figure}[h!]
\begin{center}
\includegraphics[width=16cm]{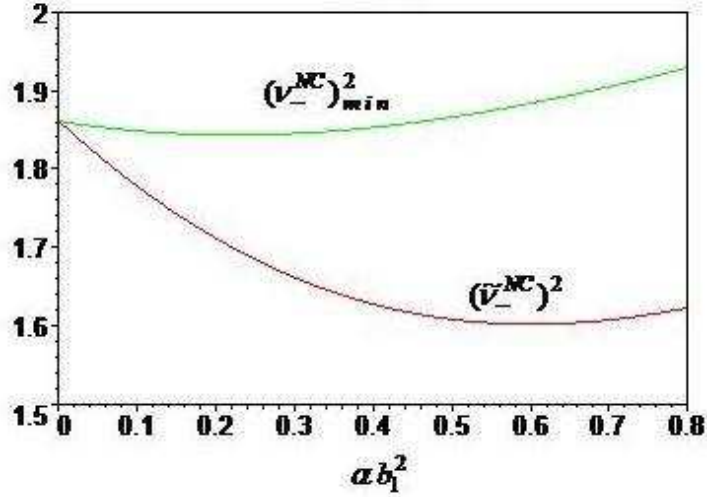}
\caption{(Coloronline)
The smallest symplectic eigenvalue $(\tilde{\nu}_{x_{-}}^{NC})^2$ of the
partially transformed variance matrix $\tilde{V}$  and
the function $(\nu_{-}^{NC}(\theta))_{min} $ representing the entanglement
criterion are plotted
versus $\alpha b_{1}^{2}$. Here $\alpha_{min} \simeq 1.4/\theta$.
The state is entangled
since $(\tilde{\nu}_{x_{-}}^{NC})^2 < (\nu_{-}^{NC}(\theta))^2_{min} $
for all values in the parameter space.}
\end{center}
\label{f3}
\end{figure}

\begin{figure}[h!]
\begin{center}
\includegraphics[width=16cm]{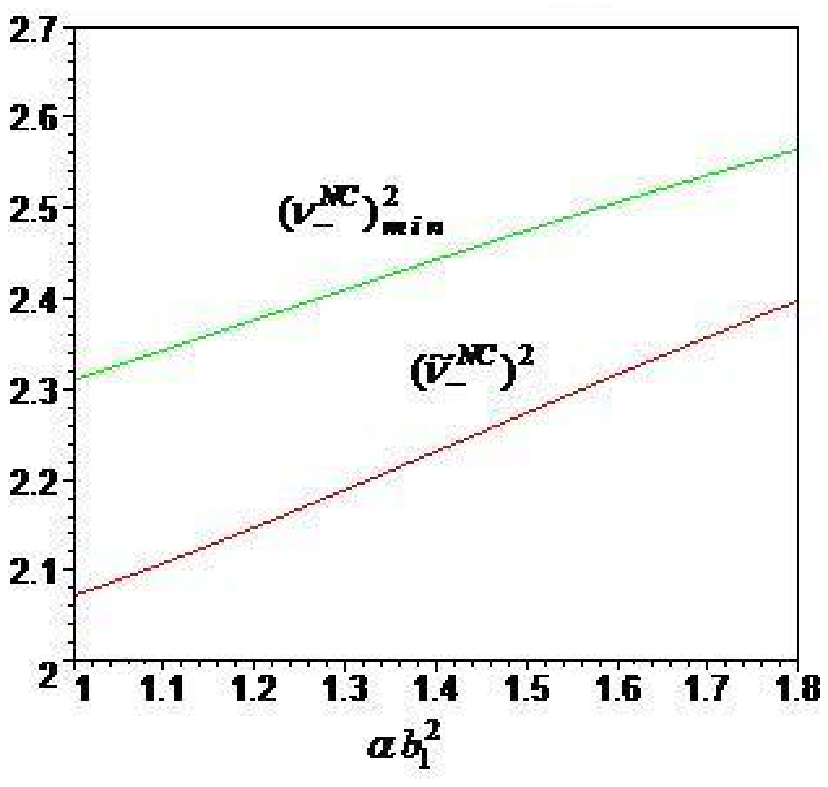}
\caption{(Coloronline)
The smallest symplectic eigenvalue $(\tilde{\nu}_{x_{-}}^{NC})^2$ of the
partially transformed variance matrix $\tilde{V}$  and
the function $(\nu_{-}^{NC} (\theta))_{min}$ representing the entanglement
criterion are plotted
versus $\alpha b_{1}^{2}$. Here $\alpha_{min} \simeq 1.6/\theta$.
The state is entangled
since  $(\tilde{\nu}_{x_{-}}^{NC})^2 < (\nu_{-}^{NC}(\theta))^2_{min} $
for all values
in the parameter space.}
\end{center}
\label{f4}
\end{figure}

\begin{figure}[h!]
\begin{center}
\includegraphics[width=16cm]{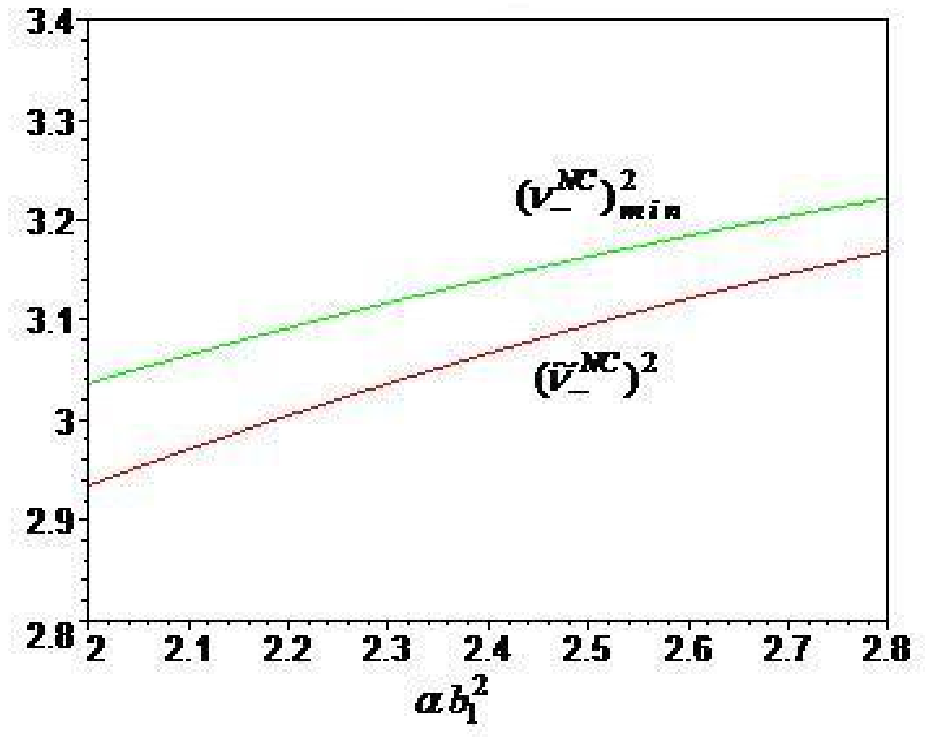}
\caption{(Coloronline)
The smallest symplectic eigenvalue $(\tilde{\nu}_{x_{-}}^{NC})^2$ of the
partially transformed variance matrix $\tilde{V}$  and
the function $(\nu_{-}^{NC} (\theta))_{min}$ representing the
entanglement criterion are plotted
versus $\alpha b_{1}^{2}$. Here $\alpha_{min} \simeq 1.8/\theta$.
The state is entangled
since  $(\tilde{\nu}_{x_{-}}^{NC})^2 < (\nu_{-}^{NC}(\theta))^2_{min} $
for all values
in the parameter space.}
\end{center}
\label{f5}
\end{figure}

Next, in order to apply the PPT entanglement criterion,
we perform partial
transposition of the variance matrix,
which corresponds to time reversal, or in
terms of continuous variables, sign change of momentum variables.
Thus, when we take the partial transpose, we have to change the
sign of all the components of the momentum variables of one of the
particles. Therefore, the variance matrix $V$ transforms under
partial transposition as
\begin{eqnarray}
V\rightarrow\tilde{V}=\Lambda V \Lambda \label{pt1}
\end{eqnarray}
where $\Lambda= diag(1,1,1,1,1,1,-1,-1)$.
After performing partial transposition operation, we compute the
eigenvalues $\tilde{\nu}_{-}$ of the matrix
$|2i\Omega \tilde{V}|$. Due to the partial transposition
operation, the symplectic invariant quantities $det\gamma_{a}$ and
$det\gamma_{b}$ just flip their sign. Hence, the expression for
the smallest symplectic eigenvalue can be written as
\begin{eqnarray}
\tilde{\nu}_{-}^{NC} = min(\tilde{\nu}_{x_{-}}^{NC},\tilde{\nu}_{y_{-}}^{NC})
\label{sse2}
\end{eqnarray}
where
\begin{eqnarray}
\tilde{\nu}_{x_{-}}^{NC} &=& \sqrt{2\Delta_{x}(\tilde{V})- 2
\sqrt{(\Delta_{x}(\tilde{V}))^{2} - 4 \det\delta_{a}}},\quad
\Delta_{x}(\tilde{V}) = \det\alpha_{a} + \det\beta_{a} - 2\det\gamma_{a}
\label{se1}, \\
\tilde{\nu}_{y_{-}}^{NC} &=& \sqrt{2\Delta_{y}(\tilde{V})- 2
\sqrt{(\Delta_{y}(\tilde{V}))^{2} - 4\det\delta_{b}}},\quad
\Delta_{y}(\tilde{V}) = \det\alpha_{b} + \det\beta_{b}-2\det\gamma_{b}
\label{se2}
\end{eqnarray}
For a non-commutative system the state is entangled when the smallest
symplectic eigenvalue $\tilde{\nu}_{-}\equiv\tilde{\nu}_{-}^{NC}$
given in Eq.(\ref{sse2}) satisfies the condition (following from
Eq.(\ref{modphys}))
\begin{eqnarray}
(\tilde{\nu_{-}}^{NC})^2 < (\nu_{-}^{NC} (\theta))^2_{min}
\label{phys.con.2}
\end{eqnarray}

We now perform the partial transposition of the variance matrix
(\ref{vm5})  and compute the symplectic eigenvalues
$\tilde{\nu}_{x_{-}}^{NC}$ and $\tilde{\nu}_{y_{-}}^{NC}$ of the
partially transposed matrix with the help of Eqs.(\ref{se1}) and
(\ref{se2}).  These are given by
\begin{eqnarray}
(\tilde{\nu}_{x_{-}}^{NC})^{2}&&=4N^{4}[1+\frac{3\theta^{2}\alpha^{2}}{4}+
\frac{4(8\theta^{4}\alpha^{5}b_{1}^{2}+3\theta^{4}\alpha^{4}+16\theta^{2}\alpha^{2}+16)exp(\frac{-8\alpha
b_{1}^{2}}{(\alpha^{2}\theta^{2}+4)})}{(\alpha^{2}\theta^{2}+4)^{3}}{}\nonumber\\&&
-\frac{256N^{2}\theta^{4}\alpha^{5}b_{1}^{2}exp(\frac{-16\alpha
b_{1}^{2}}{(\alpha^{2}\theta^{2}+4)})}{(\alpha^{2}\theta^{2}+4)^{4}}]
[1+\frac{16(\alpha^{2}\theta^{2}-16\alpha
b_{1}^{2}+4)exp(\frac{-8\alpha
b_{1}^{2}}{((\alpha^{2}\theta^{2}+4)})}{(\alpha^{2}\theta^{2}+4)^{3}}]
  \label{se3}, \\
(\tilde{\nu}_{y_{-}}^{NC})^{2}&&=4N^{4}[1+\frac{3\theta^{2}\alpha^{2}}{4}+
\frac{4(3\theta^{2}\alpha^{2}+4)exp(\frac{-8\alpha
b_{1}^{2}}{(\alpha^{2}\theta^{2}+4)})}{(\alpha^{2}\theta^{2}+4)^{2}}]
 [1+\frac{16exp(\frac{-8\alpha
b_{1}^{2}}{((\alpha^{2}\theta^{2}+4)})}{(\alpha^{2}\theta^{2}+4)^{2}}]
 \label{se4}
\end{eqnarray}
where $N^{2}=\frac{1}{2(1+ 4/5 \exp(-8\alpha b_{1}^{2}/5))}$.  The state
is entangled if the smallest symplectic eigenvalue satisfies the
condition (\ref{phys.con.2}).

The smallest symplectic eigenvalue $(\tilde{\nu}_{x_{-}}^{NC})^2$  is
 plotted versus $\alpha b_{1}^{2}$ in Figs. 3, 4 and 5 respectively for different
ranges of the parameter $b_1$ corresponding to the respective
displayed values of $\alpha_{min}$.  From the figures it can
 be seen that $\tilde{\nu}_{x_{-}}^{NC}$ satisfies the condition
 (\ref{phys.con.2}), for a range of parameter values thus
 showing that the state is entangled for these parameter
 values.

Now we are in a position to compare the amount of entanglement in the
commutative (i.e.,taking the $\theta=0$ limit in the composite state
$\Psi$ of Eq.(\ref{sai})) and non-commutative (i.e.,$~\theta\neq 0$)
case, respectively.  The respective magnitudes of entanglement can be
quantified using the logarithmic negativity\cite{logneg}.  The two
expressions can be formally written as
\begin{eqnarray}
E^{C} &=& max[0,-\frac{1}{2}log_{2}(\tilde{\nu}_{-}^{C})^2]
\label{comm.ent}, \\
E^{NC} &=& max[0,-\frac{1}{2}log_{2}(\tilde{\nu}_{-})^2] \label{non-comm.ent}
\end{eqnarray}
where $\tilde{\nu}_{-}^2 = (\tilde{\nu}_{-}^{NC})^2 - (\nu_{-}^{NC}(\theta))_{min}^2 +1$, and  the
smallest symplectic eigenvalues $\tilde{\nu}_{-}^{C}$ and
$\tilde{\nu}_{-}^{NC}$ are obtained from Eqs.(\ref{comm.sse}) and
(\ref{sse2}), respectively.

\begin{figure}[h!]
\begin{center}
\includegraphics[width=16cm]{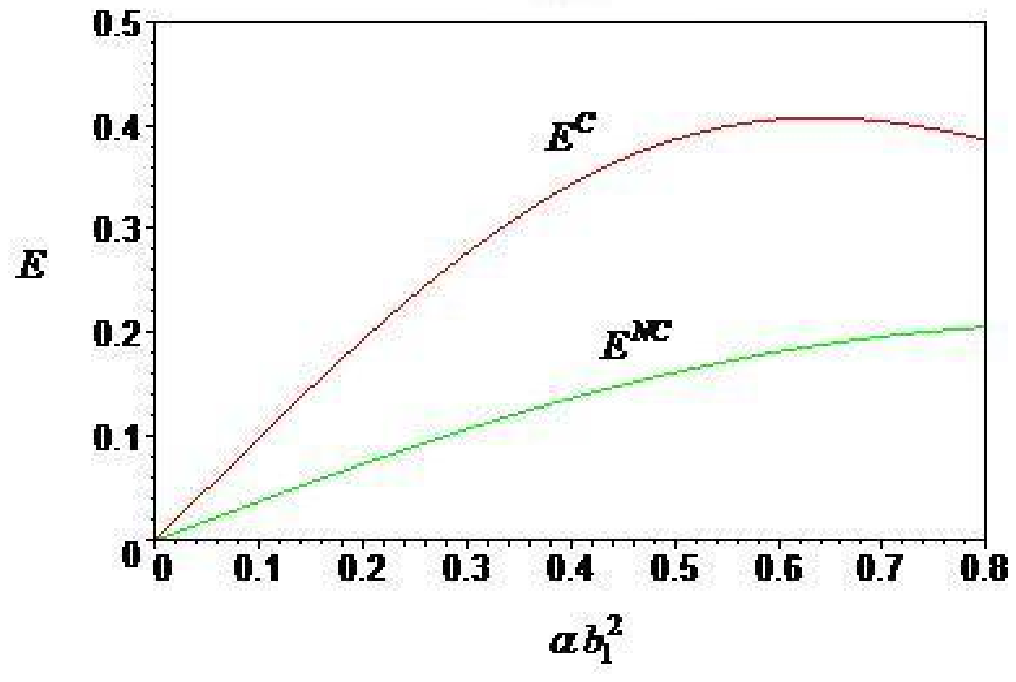}
\caption{(Coloronline) The logarithmic negativity $E^C$ and $E^{NC}$ are
plotted versus
$\alpha b_{1}^{2}$ respectively for the commutative and non-commutative systems.
Here $\alpha_{min} \simeq 1.4/\theta$.}
\end{center}
\label{f6}
\end{figure}

\begin{figure}[h!]
\begin{center}
\includegraphics[width=16cm]{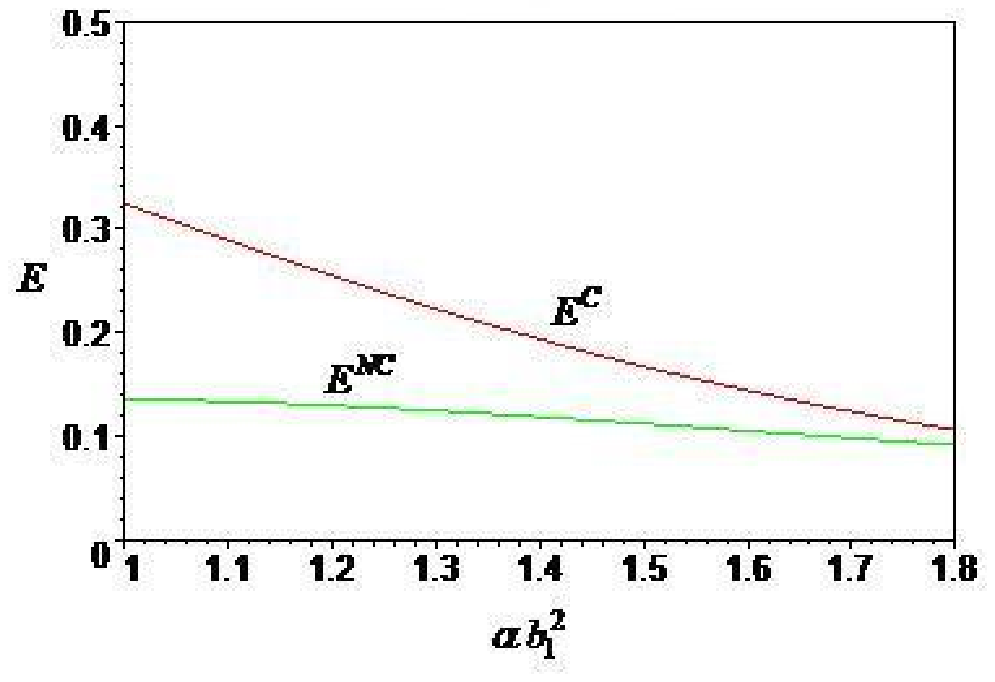}
\caption{(Coloronline) The logarithmic negativity $E^C$ and $E^{NC}$ are
plotted versus
$\alpha b_{1}^{2}$ respectively for the commutative and non-commutative systems.
Here $\alpha_{min} \simeq 1.6/\theta$.}
\end{center}
\label{f7}
\end{figure}

\begin{figure}[h!]
\begin{center}
\includegraphics[width=16cm]{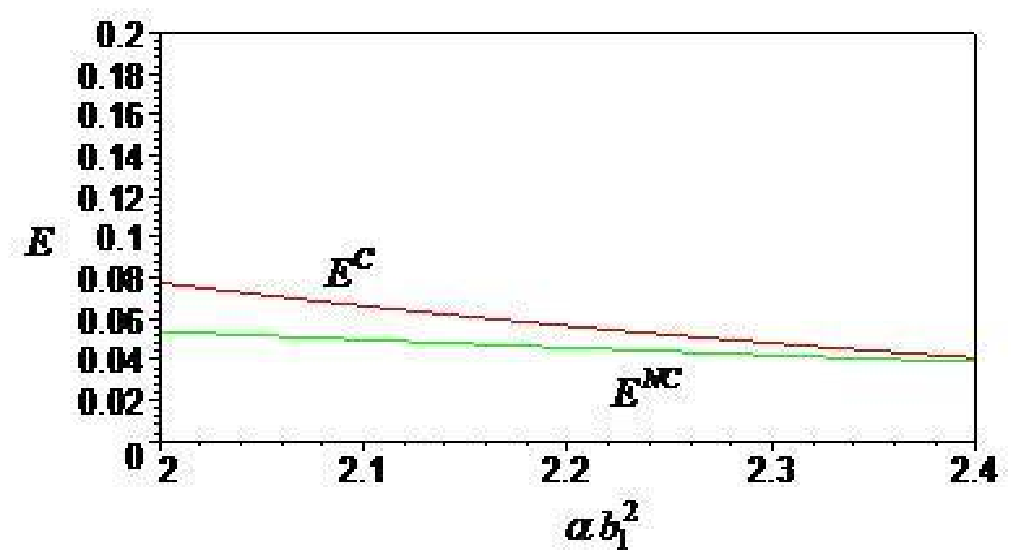}
\caption{(Coloronline) The logarithmic negativity $E^C$ and $E^{NC}$ are
plotted versus
$\alpha b_{1}^{2}$ respectively for the commutative and non-commutative systems.
Here $\alpha_{min} \simeq 1.8/\theta$.}
\end{center}
\label{f8}
\end{figure}

In Figs. 6, 7 and 8 the magnitude of entanglement $E$ is plotted versus
$\alpha b_{1}^{2}$ respectively for different ranges of $b_1$ (corresponding
to the appropriate values of $\alpha_{min}$ for the respective ranges of $b_1$)
for the commutative as well as the non-commutative case
separately. One can see that the underlying non-commutativity leads to
a reduction of entanglement in all the cases. This feature is most
prominent, i.e., the difference in the magnitute of entanglement
between the commutative and the non-commutative system is maximum, when
$\alpha b_1^2 \sim O(1)$. For larger separation of the two particles
($b_1 = a_1$ for the case we
have studied), this difference tends to decrease, and the state becomes
separable for asymptotically large values of $b_1$ in the commutative as
well as the non-commutative case. This
clear manifestation of non-commutativity hindering the entanglement of two
particles at small distances, could presumably be
correlated to the enhanced variances of the physical observables coming from the
modified space-space and phase-space uncertainty relations (\ref{spread1})
and (\ref{spread2}) in non-commutative space.

\section{Summary and Conclusions}

In this paper we have investigated the role of position-space
noncommutativity on the entanglement of bipartite systems. We focus on
Gaussian states in the continuous variable position-momentum phase
space, where the formalism for detecting and quantifying
bipartite entanglement is well-established
\cite{simon1,duan,reviews}. We have first extended the symplectic
formalism for studying entanglement for a bipartite
Gaussian state residing on a two- (spatial) dimensional plane. The
validity of this formulation is illustrated with the
help of an example of a two-particle state which is shown to exhibit
expected entanglement properties. We then prescribe our formulation
for extending the PPT criterion for
entanglement of non-commutative systems. The $\theta$-dependence
of the entanglement criterion is derived.
As an off-shoot we show that for a single particle on the two-dimensional
noncommutative plane,
whose state may be viewed as a composite state of two one-dimensional
modes,  the underlying non-commutativity does \emph{not}
impact the {\it a priori} separability of the modes.
We have finally presented an
example of a non-commutative state of two particles. This represents a
pure state and is manifestly entangled in the commutative ($\theta
=0$) limit. As a striking feature of our formalism, we are able to
show that  the magnitude of entanglement reduces under the
effect of non-commutative dynamics for a whole range of physically allowed
parameter values.

A notable consequence of this study has been our result of
how the criterion of entanglement itself is modified in the presence
of non-commutativity. Our goal in this paper has been not to
regard entanglement as a resource for performing information processing
tasks, but rather to study the impact of noncommutativity on entanglement
itself. Indeed, the role of (anti-)symmetrization associated with the
statistics of indistinguishable particles on their entanglement is
itself not yet well-understood, and any entanglement resulting from
such (anti-) symmetrization might not be amenable for use as
resource\cite{eckert}. It may be relevant to observe here that quantum
entanglement is difficult to preserve under
the interactions of the system  with the
environment.  Decoherence\cite{zurek} is an effective phenomenon
where one is practically
unable to monitor the huge number of degrees of freedom of the
environment. It remains an open question whether the loss (or
reduction) of quantum coherence  could be an intrinsic
feature of the dynamics of individual systems, instead of being a
purely effective one. There are indeed some models such as gravity
induced state vector reduction\cite{penrose} and continuous spontaneous
localization\cite{csl}, that produce quantum decoherence as an
ingredient of the basic quantum dynamics itself, though these schemes
are yet to be verified experimentally. On the other hand, the idea of
space-space noncommutativity is well-inspired at the fundamental
level, and is certainly an off-shoot of the physics in strong
gravitational fields in the presence of event horizons of black
holes\cite{dfr}. Non-commutative dynamics being able to restrict the
entanglement of
two particles at  short distances is a remarkable consequence of
the analysis presented in this paper.
Such an effect of non-commutativity is expected to have interesting
consequences on the entanglement of quantum states and related
information capacities wherever position-space noncommutativity is
effective.

\vskip 0.2in

{\it Acknowledgments:} We would like to thank Subhash Chaturvedi and
Frederik G. Scholtz
for discussions, and Soumen Mandal for help
using {\it Mathematica} to compute the integrals for the
non-commutative variance matrix elements. ASM would like to acknowledge
support from a project funded by DST, India.

\subsection*{Appendix I: Standard form of the variance matrix}

By the following steps one can bring a variance matrix into the
standard form from which its symplectic eigen values could be easily
obtained.  Step-I: The variance matrix can be expressed in the basis
$(x_{1}^{(1)},p_{1}^{(1)},x_{2}^{(1)},p_{2}^{(1)},x_{1}^{(2)},p_{1}^{(2)},
,x_{2}^{(2)},p_{2}^{(2)})$ in terms of $2\times2$ block matrices as:\\
\begin{eqnarray}
\textsl{V} = \left(\begin{matrix}{\alpha_{x}^{(1)} &
 \alpha_{x,y}^{(1)} & \gamma_{x}^{(1,2)} &
 \gamma_{x,y}^{(1,2)} \cr
 \alpha_{x,y}^{(1)}& \alpha_{y}^{(1)} &
 (\gamma_{x,y}^{(1,2)})^{T} &
 \gamma_{y}^{(1,2)} \cr
(\gamma_{x}^{(1,2)})^{T} &\gamma_{x,y}^{(1,2)}
&\beta_{x}^{(2)} & \beta_{x,y}^{(2)}
 \cr (\gamma_{x,y}^{(1,2)})^{T} & (\gamma_{y}^{(1,2)})^{T}
 & \beta_{x,y}^{(2)} &
 \beta_{y}^{(2)}
 }\end{matrix}\right)_{8\times8}
\label{newbasis}
\end{eqnarray}
The superscripts $1$ and $2$ refer to the first and second particle,
respectively, and subscripts $x_{1}$ and $x_{2}$ refer to the two
perpendicular axes $x$ and $y$, respectively.  Using Williamson's
theorem \cite{william}, we can choose two local symplectic
transformations $S_{1},S_{2} \in Sp(4,R)$ which transform
\begin{eqnarray}
V \rightarrow (S_{1}\otimes S_{2}) V (S_{1}^{T}\otimes
S_{2}^{T})=V^{\prime}\equiv \left(\begin{matrix}{g_{a}I &
 0 & (\gamma_{x}^{(1,2)})^{\prime} &
 (\gamma_{x,y}^{(1,2)})^{\prime} \cr
 0 & g_{b}I & ((\gamma_{x,y}^{(1,2)})^{\prime})^{T} &
 (\gamma_{y}^{(1,2)})^{\prime} \cr
((\gamma_{x}^{(1,2)})^{\prime})^{T}
&(\gamma_{x,y}^{(1,2)})^{\prime} & g_{c}I & 0
 \cr ((\gamma_{x,y}^{(1,2)})^{\prime})^{T} &
 ((\gamma_{y}^{(1,2)})^{\prime})^{T} & 0& g_{d}I
 }\end{matrix}\right)
\label{transform1}
\end{eqnarray}
where $I$ denotes the $2\times2$ identity matrix and $0$ denotes the
$2\times2$ null matrix.  As a result of the transformation, the
cross-correlation block matrices are modified, and they are denoted
with prime.

Step-II: There exist symplectic transformations $S_3,S_4\in Sp(2,R)$
which transform the variance matrix $V^{\prime}$ as:
\begin{eqnarray}
V^{\prime}\rightarrow \left(\begin{matrix}{S_{3} &
 0 & 0 & 0 \cr
 0 & S_{4} & 0 & 0 \cr 0 & 0 & S_{3} & 0
 \cr 0 & 0 & 0 &  S_{4}
 }\end{matrix}\right)V^{\prime}\left(\begin{matrix}{S_{3}^{T} &
 0 & 0 & 0 \cr
 0 & S_{4}^{T} & 0 & 0 \cr 0 & 0 & S_{3}^{T} & 0
 \cr 0 & 0 & 0 &  S_{4}^{T} }\end{matrix}\right)
\end{eqnarray}
The transformation brings the $2\times2$ block matrices into
diagonal form as follows:
\begin{eqnarray}
&&(\gamma_{x}^{(1,2)})^{\prime}\rightarrow
S_{3}(\gamma_{x}^{(1,2)})^{\prime}S_{3}^{T}=\left(\begin{matrix}{m_{a}
& 0 \cr 0 & m_{c}}\end{matrix}\right){}\nonumber\\&&
(\gamma_{x,y}^{(1,2)})^{\prime}\rightarrow
S_{3}(\gamma_{x,y}^{(1,2)})^{\prime}S_{4}^{T} =
\left(\begin{matrix}{q_{a}
& 0 \cr 0 & q_{b}}\end{matrix}\right){}\nonumber\\&&
(\gamma_{y}^{(1,2)})^{\prime}\rightarrow
S_{4}(\gamma_{y}^{(1,2)})^{\prime}S_{4}^{T} = \left(\begin{matrix}{m_{b}
& 0 \cr 0 & m_{d}}\end{matrix}\right) \label{transform2}
\end{eqnarray}
Further, the transformation does not effect the $2\times2$ block
matrices of the type $\left(\begin{matrix}{g_{i}& 0 \cr 0 & g_{i}
}\end{matrix}\right)~(i=a,b,c,d)$ since they are proportional to the
identity matrix.

After performing these steps, the standard form of the variance matrix
can be expressed in the form
\begin{eqnarray}
\textsl{V} = \left(\begin{matrix}{g_{a} & 0 & 0 & 0& m_{a}& 0 &
q_{a}& 0 \cr 0 & g_{a} & 0 & 0 & 0 & m_{c} & 0 & q_{b} \cr 0 & 0 &
g_{b} & 0& q_{a} & 0 & m_{b} &0  \cr 0 & 0 & 0 & g_{b}& 0 & q_{b}
& 0 & m_{d} \cr m_{a} & 0 & q_{a} & 0 & g_{c} & 0 & 0 & 0 \cr 0 &
m_{c} & 0 & q_{b} & 0 & g_{c} & 0 & 0 \cr q_{a} & 0 & m_{b} & 0 &
0 & 0 & g_{d} & 0 \cr 0 & q_{b}& 0 & m_{d} & 0 & 0 & 0 & g_{d}
 }\end{matrix}\right)_{8\times8}
\label{twostandard}
\end{eqnarray}

\subsection*{Appendix II: Non-Commutative variance matrix elements}

The entries of the variance matrix $\overline{V}$ for the
non-commutative case are obtained from computing the following
relevant expectation values:

\begin{eqnarray}
&&\langle \hat{F}_{\theta}^{-1}(\frac{\hat{\bar{x}}_{i} \hat{\bar{x}}_{j} +
  \hat{\bar{x}}_{j}\hat{\bar{x}}_{i}}{2} \otimes I)\hat{F}_{\theta}\rangle_{\Psi} =
  N^{2}[\langle \frac{\hat{\bar{x}}_{i}\hat{\bar{x}}_{j} +
  \hat{\bar{x}}_{j}\hat{\bar{x}}_{i}}{2} \rangle_{\psi_{1}} + \langle
\frac{\hat{\bar{x}}_{i}\hat{\bar{x}}_{j} +
  \hat{\bar{x}}_{j}\hat{\bar{x}}_{i}}{2}\rangle_{\psi_{2}} +{}\nonumber\\
&&\frac{\theta}{2}
(\langle\hat{\bar{x}}_{i}\rangle_{\psi_{1}} \langle\epsilon_{jl}\hat{p}_{l}
  \rangle_{\psi_{2}} + \langle\hat{\bar{x}}_{j} \rangle_{\psi_{1}} \langle
  \epsilon_{ik}\hat{p}_{k} \rangle_{\psi_{2}} - \langle\epsilon_{jk}
  \hat{p}_{k}\rangle_{\psi_{1}} \langle\hat{\bar{x}}_{i}\rangle_{\psi_{2}}
-\langle\epsilon_{ik}\hat{p}_{k} \rangle_{\psi_{1}} \langle\hat{\bar{x}}_{j}
  \rangle_{\psi_{2}}) {}\nonumber\\
&&+\frac{\theta^{2}}{4}(\langle\epsilon_{ik}\hat{p}_{k} \epsilon_{jl}\hat{p}_{l}
  \rangle_{\psi_{1}}
+ \langle\epsilon_{ik}\hat{p}_{k}\epsilon_{jl} \hat{p}_{l}\rangle_{\psi_{2}}) +
2Re\langle(\frac{\hat{\bar{x}}_{i}\hat{\bar{x}}_{j} + \hat{\bar{x}}_{j}\hat{\bar{x}}_{i}}{2}\otimes
I)\hat{F}_{\theta}^{-2}\hat{\tau}_{0} \rangle_{\psi_{1} \otimes
  \psi_{2}}{}\nonumber\\
&&+\frac{\theta^{2}}{2} Re\langle(I\otimes \epsilon_{ik}\hat{p}_{k}\epsilon_{jl}\hat{p}_{l}
)\hat{F}_{\theta}^{-2}\hat{\tau}_{0}\rangle_{\psi_{1}\otimes\psi_{2}}+\theta
Re\langle(\hat{\bar{x}}_{i}\otimes\epsilon_{jl}\hat{p}_{l})
  \hat{F}_{\theta}^{-2}\hat{\tau}_{0}\rangle_{\psi_{1}\otimes\psi_{2}}
{}\nonumber\\
&&+\theta Re\langle(\hat{\bar{x}}_{j}\otimes \epsilon_{ik}\hat{p}_{k})
  \hat{F}_{\theta}^{-2} \hat{\tau}_{0}\rangle_{\psi_{1}\otimes\psi_{2}}],
\end{eqnarray}
\begin{eqnarray}
&&\langle
\hat{F}_{\theta}^{-1}(\frac{\hat{\bar{x}}_{i}\hat{p}_{j}+\hat{p}_{j}\hat{\bar{x}}_{i}}{2}\otimes
I)\hat{F}_{\theta}\rangle_{\Psi}=N^{2}[\langle
\frac{\hat{\bar{x}}_{i}\hat{p}_{j}+\hat{p}_{j}\hat{\bar{x}}_{i}}{2}\rangle_{\psi_{1}}+\langle
\frac{\hat{\bar{x}}_{i}\hat{p}_{j} + \hat{p}_{j}\hat{\bar{x}}_{i}}{2} \rangle_{\psi_{2}}
  +{}\nonumber\\
&&\frac{\theta}{2} (\langle \hat{p}_{j}\rangle_{\psi_{1}}\langle
  \epsilon_{ik} \hat{p}_{k}\rangle_{\psi_{2}} - \langle\epsilon_{ik}\hat{p}_{k}
  \rangle_{\psi_{1}}\langle \hat{p}_{j}\rangle_{\psi_{2}})
+2Re\langle(\frac{\hat{\bar{x}}_{i}\hat{p}_{j} + \hat{p}_{j}\hat{\bar{x}}_{i}}{2}\otimes
I)\hat{F}_{\theta}^{-2}\hat{\tau}_{0} \rangle_{\psi_{1} \otimes
  \psi_{2}}{}\nonumber\\
&&+\theta Re\langle(\hat{p}_{j}\otimes
\epsilon_{il}\hat{p}_{l})\hat{F}_{\theta}^{-2}\hat{\tau}_{0}\rangle_{\psi_{1}\otimes\psi_{2}}]
\end{eqnarray}
\begin{eqnarray}
&&\langle \hat{F}_{\theta}^{-1}(\hat{p}_{i}\hat{p}_{j}\otimes
I)\hat{F}_{\theta}\rangle_{\Psi}=N^{2}[\langle
\hat{p}_{i}\hat{p}_{j}\rangle_{\psi_{1}}+\langle \hat{p}_{i}\hat{p}_{j}\rangle_{\psi_{2}}
{}\nonumber\\&&+ 2Re\langle(\hat{p}_{i}\hat{p}_{j}\otimes
I)\hat{F}_{\theta}^{-2}\hat{\tau}_{0}\rangle_{\psi_{1}\otimes\psi_{2}}]
\end{eqnarray}
\begin{eqnarray}
&&\langle
\hat{F}_{\theta}^{-1}(I\otimes\frac{\hat{\bar{x}}_{i}\hat{\bar{x}}_{j}+\hat{\bar{x}}_{j}\hat{\bar{x}}_{i}}{2})\hat{F}_{\theta}\rangle_{\Psi}=N^{2}[\langle
\frac{\hat{\bar{x}}_{i}\hat{\bar{x}}_{j}+\hat{\bar{x}}_{j}\hat{\bar{x}}_{i}}{2}\rangle_{\psi_{1}}+\langle
\frac{\hat{\bar{x}}_{i}\hat{\bar{x}}_{j}+\hat{\bar{x}}_{j}\hat{\bar{x}}_{i}}{2}\rangle_{\psi_{2}}+{}\nonumber\\&&\frac{\theta}{2}
(\langle\hat{\bar{x}}_{i}\rangle_{\psi_{1}}\langle\epsilon_{jl}\hat{p}_{l}\rangle_{\psi_{2}}+
\langle\hat{\bar{x}}_{j}\rangle_{\psi_{1}}\langle\epsilon_{ik}\hat{p}_{k}\rangle_{\psi_{2}}
-\langle\epsilon_{jk}\hat{p}_{k}\rangle_{\psi_{1}}\langle\hat{\bar{x}}_{i}\rangle_{\psi_{2}}
-\langle\epsilon_{ik}\hat{p}_{k}\rangle_{\psi_{1}}\langle\hat{\bar{x}}_{j}\rangle_{\psi_{2}})
{}\nonumber\\&&+\frac{\theta^{2}}{4}(\langle\epsilon_{ik}\hat{p}_{k}\epsilon_{jl}\hat{p}_{l}\rangle_{\psi_{1}}
+\langle\epsilon_{ik}\hat{p}_{k}\epsilon_{jl}\hat{p}_{l}\rangle_{\psi_{2}})+
2Re\langle(I\otimes\frac{\hat{\bar{x}}_{i}\hat{\bar{x}}_{j}+\hat{\bar{x}}_{j}\hat{\bar{x}}_{i}}{2})
\hat{F}_{\theta}^{-2}\hat{\tau}_{0}\rangle_{\psi_{1}\otimes\psi_{2}}{}\nonumber\\&&+\frac{\theta^{2}}{2}
Re\langle(\epsilon_{ik}\hat{p}_{k}\epsilon_{jl}\hat{p}_{l}\otimes I
)\hat{F}_{\theta}^{-2}\hat{\tau}_{0}\rangle_{\psi_{1}\otimes\psi_{2}}-\theta
Re\langle(\epsilon_{ik}\hat{p}_{k}\otimes\hat{\bar{x}}_{j})\hat{F}_{\theta}^{-2}\hat{\tau}_{0}\rangle_{\psi_{1}\otimes\psi_{2}}
{}\nonumber\\&&-\theta
Re\langle(\epsilon_{jl}\hat{p}_{l}\otimes\hat{\bar{x}}_{i})\hat{F}_{\theta}^{-2}\hat{\tau}_{0}\rangle_{\psi_{1}\otimes\psi_{2}}]
\end{eqnarray}
\begin{eqnarray}
&&\langle
\hat{F}_{\theta}^{-1}(I\otimes\frac{\hat{\bar{x}}_{i}\hat{p}_{j}+\hat{p}_{j}\hat{\bar{x}}_{i}}{2}
)\hat{F}_{\theta}\rangle_{\Psi}=N^{2}[\langle
\frac{\hat{\bar{x}}_{i}\hat{p}_{j}+\hat{p}_{j}\hat{\bar{x}}_{i}}{2}\rangle_{\psi_{1}}+\langle
\frac{\hat{\bar{x}}_{i}\hat{p}_{j}+\hat{p}_{j}\hat{\bar{x}}_{i}}{2}\rangle_{\psi_{2}}+{}\nonumber\\&&\frac{\theta}{2}
(\langle
\hat{p}_{j}\rangle_{\psi_{1}}\langle\epsilon_{ik}\hat{p}_{k}\rangle_{\psi_{2}}-
\langle\epsilon_{ik}\hat{p}_{k}\rangle_{\psi_{1}}\langle
\hat{p}_{j}\rangle_{\psi_{2}})
+2Re\langle(I\otimes\frac{\hat{\bar{x}}_{i}\hat{p}_{j}+\hat{p}_{j}\hat{\bar{x}}_{i}}{2})\hat{F}_{\theta}^{-2}\hat{\tau}_{0}\rangle_{\psi_{1}\otimes\psi_{2}}
{}\nonumber\\&&-\theta Re\langle(\epsilon_{ik}\hat{p}_{k}\otimes
\hat{p}_{j})\hat{F}_{\theta}^{-2}\hat{\tau}_{0}\rangle_{\psi_{1}\otimes\psi_{2}}]
\end{eqnarray}
\begin{eqnarray}
&&\langle \hat{F}_{\theta}^{-1}(I\otimes
\hat{p}_{i}\hat{p}_{j})\hat{F}_{\theta}\rangle_{\Psi}=N^{2}[\langle
\hat{p}_{i}\hat{p}_{j}\rangle_{\psi_{1}}+\langle \hat{p}_{i}\hat{p}_{j}\rangle_{\psi_{2}}
{}\nonumber\\&&+ 2Re\langle(I\otimes
\hat{p}_{i}\hat{p}_{j})\hat{F}_{\theta}^{-2}\hat{\tau}_{0}\rangle_{\psi_{1}\otimes\psi_{2}}]
\end{eqnarray}
\begin{eqnarray}
&&\langle \hat{F}_{\theta}^{-1}(\hat{\bar{x}}_{i}\otimes
\hat{\bar{x}}_{j})\hat{F}_{\theta}\rangle_{\Psi}=N^{2}[\langle
\hat{\bar{x}}_{i}\rangle_{\psi_{1}}\langle
\hat{\bar{x}}_{j}\rangle_{\psi_{2}}+\langle
\hat{\bar{x}}_{j}\rangle_{\psi_{1}}\langle
\hat{\bar{x}}_{i}\rangle_{\psi_{2}}+\frac{\theta}{2}(-\langle
\hat{\bar{x}}_{i}\epsilon_{jk}\hat{p}_{k}\rangle_{\psi_{1}}-{}\nonumber\\&&\langle
\hat{\bar{x}}_{j}\epsilon_{ik}\hat{p}_{k}\rangle_{\psi_{1}})+\langle
\hat{\bar{x}}_{i}\epsilon_{jk}\hat{p}_{k}\rangle_{\psi_{2}}+\langle
\hat{\bar{x}}_{j}\epsilon_{ik}\hat{p}_{k}\rangle_{\psi_{2}})+\frac{\theta^{2}}{4}(\langle
\hat{p}_{i}\rangle_{\psi_{1}}\langle \hat{p}_{j}\rangle_{\psi_{2}}-\langle
\hat{p}_{j}\rangle_{\psi_{1}}\langle
\hat{p}_{i}\rangle_{\psi_{2}}){}\nonumber\\&& +
2Re\langle(\hat{\bar{x}}_{i}\otimes
\hat{\bar{x}}_{j})\hat{F}_{\theta}^{-2}\hat{\tau}_{0}\rangle_{\psi_{1}\otimes\psi_{2}} +
\theta Re\langle(I\otimes
\epsilon_{ik}\hat{p}_{k}\hat{\bar{x}}_{j})\hat{F}_{\theta}^{-2}\hat{\tau}_{0}\rangle_{\psi_{1}\otimes\psi_{2}}
{}\nonumber\\&& - \theta Re\langle(
\hat{\bar{x}}_{i}\epsilon_{jk}\hat{p}_{k}\otimes I
)\hat{F}_{\theta}^{-2}\hat{\tau}_{0}\rangle_{\psi_{1}\otimes\psi_{2}}+\frac{\theta^{2}}{2}
Re\langle(\hat{p}_{i}\otimes \hat{p}_{j})
\hat{F}_{\theta}^{-2}\hat{\tau}_{0}\rangle_{\psi_{1}\otimes\psi_{2}}]
\end{eqnarray}
\begin{eqnarray}
&&\langle \hat{F}_{\theta}^{-1}(\hat{\bar{x}}_{i}\otimes
\hat{p}_{j})\hat{F}_{\theta}\rangle_{\Psi}=N^{2}[\langle
\hat{\bar{x}}_{i}\rangle_{\psi_{1}}\langle
\hat{p}_{j}\rangle_{\psi_{2}}+\langle \hat{p}_{j}\rangle_{\psi_{1}}\langle
\hat{\bar{x}}_{i}\rangle_{\psi_{2}}-\frac{\theta}{2}(\langle
\epsilon_{ik}\hat{p}_{k}\hat{p}_{j}\rangle_{\psi_{1}}-{}\nonumber\\&&\langle
\epsilon_{ik}\hat{p}_{k}\hat{p}_{j}\rangle_{\psi_{2}})+
2Re\langle(\hat{\bar{x}}_{i}\otimes
\hat{p}_{j})\hat{F}_{\theta}^{-2}\hat{\tau}_{0}\rangle_{\psi_{1}\otimes\psi_{2}}+
\theta Re\langle(I\otimes
\epsilon_{ik}\hat{p}_{k}\hat{p}_{j})\hat{F}_{\theta}^{-2}\hat{\tau}_{0}\rangle_{\psi_{1}\otimes\psi_{2}}]
\end{eqnarray}
\begin{eqnarray}
&&\langle \hat{F}_{\theta}^{-1}(\hat{p}_{i}\otimes
\hat{\bar{x}}_{j})\hat{F}_{\theta}\rangle_{\Psi}=N^{2}[\langle
\hat{\bar{x}}_{j}\rangle_{\psi_{1}}\langle
\hat{p}_{i}\rangle_{\psi_{2}}+\langle \hat{p}_{i}\rangle_{\psi_{1}}\langle
\hat{\bar{x}}_{j}\rangle_{\psi_{2}}-\frac{\theta}{2}(\langle
\epsilon_{jk}\hat{p}_{k}\hat{p}_{i}\rangle_{\psi_{1}}-{}\nonumber\\&&\langle
\epsilon_{jk}\hat{p}_{k}\hat{p}_{i}\rangle_{\psi_{2}})+
2Re\langle(\hat{p}_{i}\otimes
\hat{\bar{x}}_{j})\hat{F}_{\theta}^{-2}\hat{\tau}_{0}\rangle_{\psi_{1}\otimes\psi_{2}}-
\theta Re\langle( \epsilon_{jk}\hat{p}_{k}\hat{p}_{i}\otimes I
)\hat{F}_{\theta}^{-2}\hat{\tau}_{0}\rangle_{\psi_{1}\otimes\psi_{2}}]
\end{eqnarray}
\begin{eqnarray}
&&\langle \hat{F}_{\theta}^{-1}(\hat{p}_{i}\otimes
\hat{p}_{j})\hat{F}_{\theta}\rangle_{\Psi}=N^{2}[\langle
\hat{p}_{i}\rangle_{\psi_{1}}\langle \hat{p}_{j}\rangle_{\psi_{2}}+\langle
\hat{p}_{j}\rangle_{\psi_{1}}\langle
\hat{p}_{i}\rangle_{\psi_{2}}{}\nonumber\\&&+ 2Re\langle(\hat{p}_{i}\otimes
\hat{p}_{j})\hat{F}_{\theta}^{-2}\hat{\tau}_{0}\rangle_{\psi_{1}\otimes\psi_{2}}]
\end{eqnarray}

Using the above relations the elements of the CM given by Eq.(\ref{vm4})
are obtained to be
\begin{eqnarray}
\bar{A}_{ij} &=& \langle \hat{F}_{\theta}^{-1}(\frac{(\hat{\overline{x_{i}}}~
\hat{\overline{x_{j}}} + \hat{\overline{x_{j}}}~\hat{\overline{x_{i}}})}{2}\otimes
I)\hat{F}_{\theta}\rangle_{\Psi}-\langle \hat{F}_{\theta}^{-1}(\hat{\overline{x}_{i}}\otimes
I)\hat{F}_{\theta}\rangle_{\Psi}\langle \hat{F}_{\theta}^{-1}(\hat{\overline{x}_{j}}\otimes
I)\hat{F}_{\theta}\rangle_{\Psi}{}\nonumber\\
&=& \langle \hat{F}_{\theta}^{-1}(I \otimes
\frac{(\hat{\overline{x_{i}}}\hat{\overline{x_{j}}} +
 \hat{ \overline{x_{j}}}\hat{\overline{x_{i}}})}{2}) \hat{F}_{\theta}\rangle_{\Psi} -
\langle \hat{F}_{\theta}^{-1}(I\otimes
\hat{\overline{x}_{i}})\hat{F}_{\theta}\rangle_{\Psi}\langle \hat{F}_{\theta}^{-1}(I\otimes
\hat{\overline{x}_{j}})\hat{F}_{\theta}\rangle_{\Psi}{}\nonumber\\
&=& N^{2}[\frac{\theta^{2}}{4}(\epsilon_{ik}p_{0k}\epsilon_{jl}p_{0l}
+ 2\alpha\delta_{ij}) - \frac{\theta}{2}(\epsilon_{jk}p_{0k}b_{i} +
\epsilon_{ik}p_{0k}b_{j}) + 2b_{i}b_{j} + \frac{\delta_{ij}}{\alpha} +
    {}\nonumber\\
&&2 \exp(\frac{-(p_{01}^2 + p_{02}^2)}{2\alpha})
    \exp(\frac{-8\alpha(b_{1}^2 + b_{2}^2)}{(\alpha^2\theta^2 +
      4)})[\{\theta^{4}\alpha^{4}(-p_{0i}p_{0j} +
      4\alpha^{2}b_{i}b_{j} + 4\alpha\delta_{ij})-{} \nonumber\\
&&4\theta^{3}\alpha^{4}(\epsilon_{jk}b_{k}p_{0i} +
      \epsilon_{ik}b_{k}p_{0j}) - 8\theta^{2} \alpha^{2}(p_{0i}p_{0j} +
      2\alpha^{2}\epsilon_{ik} b_{k}\epsilon_{jl} b_{l}-3\alpha
      \delta_{ij})- {}\nonumber\\
&&16\alpha^{2}\theta(\epsilon_{jk}b_{k}p_{0i} +
      \epsilon_{ik}b_{k}p_{0j}) - 16p_{0i}p_{0j} + 32\alpha\delta_{ij}\}
\{\alpha^{-2}(\alpha^2 \theta^2 + 4)^{-3}\}]-{}\nonumber\\
&&N^{2}(\frac{-\theta}{2}\epsilon_{ik}p_{0k} + \frac{8\alpha^{2}
      \theta^{2}b_{i}}{(\alpha^{2}\theta^{2}+4)^{2}}
    \exp(\frac{-(p_{01}^2 + p_{02}^2)}{2\alpha})
    \exp(\frac{-8\alpha (b_{1}^2 + b_{2}^2)}{(\alpha^2\theta^2 +
      4)}))\times{}\nonumber\\
&&(\frac{-\theta}{2}\epsilon_{jk}p_{0k} + \frac{8\alpha^{2}\theta^{2}b_{j}}
{(\alpha^{2}\theta^{2} + 4)^{2}}\exp(\frac{-(p_{01}^2 + p_{02}^2)}{2\alpha})
\exp(\frac{-8\alpha(b_{1}^2+b_{2}^2)}{(\alpha^2\theta^2 + 4)}))], \\
\bar{B}_{ij} &=& \langle \hat{F}_{\theta}^{-1} (\frac{(\hat{\overline{x_{i}}}\hat{p}_{j}
  + \hat{p}_{j}\hat{\overline{x_{i}}})}{2}\otimes I)\hat{F}_{\theta}\rangle_{\Psi} - \langle
\hat{F}_{\theta}^{-1}(\hat{\overline{x}}_{i} \otimes I)\hat{F}_{\theta}\rangle_{\Psi}
\langle \hat{p}_{j}\otimes I\rangle_{\Psi}{}\nonumber\\
&=&\langle \hat{F}_{\theta}^{-1}(I \otimes \frac{(\hat{\overline{x_{i}}}\hat{p}_{j} +
  \hat{p}_{j}\hat{\overline{x_{i}}})}{2}) \hat{F}_{\theta}\rangle_{\Psi}  - \langle
\hat{F}_{\theta}^{-1}(I\otimes \hat{\overline{x}}_{i})\hat{F}_{\theta} \rangle_{\Psi}
\langle I \otimes \hat{p}_{j}\rangle_{\Psi}{}\nonumber\\
&=&N^{2}[\frac{-\theta}{4}p_{0j} \epsilon_{ik}p_{0k} -\frac{\theta}{2}
  \alpha\epsilon_{ij}+p_{0j}b_{i} + 8 \exp(\frac{-(p_{01}^2 +
    p_{02}^2)}{2 \alpha}) \exp(\frac{-8\alpha(b_{1}^2 +
    b_{2}^2)}{(\alpha^2\theta^2 + 4)})\times{}\nonumber\\
&& [\{\theta^{3}\alpha^{3}(2\alpha\epsilon_{jk}b_{k}b_{i} -
    \epsilon_{ij}) + 2\theta^{2}\alpha^{2}p_{0i}b_{j} + 4\theta\alpha(2\alpha
b_{j}\epsilon_{ik}b_{k} -
\epsilon_{ij})+8p_{0i}b_{j}\}\times{}\nonumber\\
&&\{(\alpha^2\theta^2+4)^{-3}\}] - N^{2}(\frac{-\theta}{2}
  \epsilon_{ik}p_{0k} + \frac{8\alpha^{2}\theta^{2}b_{i}} {(\alpha^{2}
    \theta^{2} + 4)^{2}}exp(\frac{-(p_{01}^2 + p_{02}^2)}{2\alpha})
  \exp(\frac{-8 \alpha(b_{1}^2 + b_{2}^2)}{(\alpha^2\theta^2 + 4)})){}
  \nonumber\\
&&\times(+\frac{16\alpha^{2}\theta \epsilon_{jk} b_{k}}{(\alpha^{2}
    \theta^{2} + 4)^{2}} \exp(\frac{-(p_{01}^2 + p_{02}^2)}{2\alpha})
  \exp(\frac{-8\alpha(b_{1}^2 + b_{2}^2)}{(\alpha^2\theta^2+4)}))], \\
\bar{C}_{ij} &=& \langle \hat{F}_{\theta}^{-1}(\hat{\overline{x_{i}}}\otimes
\hat{\overline{x_{j}}})\hat{F}_{\theta}\rangle_{\Psi}-\langle
\hat{F}_{\theta}^{-1}(\hat{\overline{x}}_{i}\otimes I)\hat{F}_{\theta}\rangle_{\Psi}
\langle \hat{F}_{\theta}^{-1}(I\otimes \hat{\overline{x}}_{j})
\hat{F}_{\theta}\rangle_{\Psi} {}\nonumber\\
&=& N^{2}[\frac{\theta}{2}(b_{j} \epsilon_{ik}p_{0k} +
  b_{i}\epsilon_{jk} p_{0k})-2b_{i}b_{j} + 2 \exp(\frac{-(p_{01}^2 +
    p_{02}^2)}{2 \alpha})\times{}\nonumber\\
&&\exp(\frac{-8\alpha(b_{1}^2 + b_{2}^2)}{(\alpha^2\theta^2 + 4)})
[\{\theta^{4}\alpha^{4}(p_{0i}p_{0j} + 4\alpha^{2}b_{i} b_{j}) +
  4\theta^{3}\alpha^{4}b_{k} (\epsilon_{ik}p_{0j} +
  \epsilon_{jk}p_{0i}){}\nonumber\\
&+& 8\theta^{2}\alpha^{2}(p_{0i}p_{0j}(1+2\alpha^{2}) + 2\alpha^{2}
  \epsilon_{ik}\epsilon_{jl} b_{k} b_{l}) + 16\theta\alpha^{2}
  b_{k}(\epsilon_{ik}p_{0j} + \epsilon_{jk}p_{0i})+16p_{0i}p_{0j}\}
{}\nonumber\\
&&\{\alpha^{-2}(\alpha^2\theta^2 + 4)^{-3}\}]-N^{2}(\frac{-\theta}{2}
\epsilon_{ik}p_{0k} + \frac{8\alpha^{2}\theta^{2}b_{i}} {(\alpha^{2}
  \theta^{2} + 4)^{2}}\exp(\frac{-(p_{01}^2 + p_{02}^2)}{2\alpha})
\times \nonumber \\
&&\times{} \exp(\frac{-8\alpha(b_{1}^2 + b_{2}^2)}{(\alpha^2\theta^2 +
  4)}))(\frac{-\theta}{2}\epsilon_{jk} p_{0k} + \frac{8\alpha^{2}
  \theta^{2} b_{j}}{(\alpha^{2} \theta^{2}+4)^{2}}
\exp(\frac{-(p_{01}^2 + p_{02}^2)}{2\alpha})
 \exp(\frac{-8\alpha(b_{1}^2+b_{2}^2)}{(\alpha^2\theta^2+4)}))], \\
\bar{D}_{ij} &=& \langle \hat{F}_{\theta}^{-1}(\hat{\overline{x_{i}}}\otimes
\hat{p}_{j})\hat{F}_{\theta}\rangle_{\Psi}-\langle \hat{F}_{\theta}^{-1} (\hat{\overline{x}}_{i}\otimes
I)\hat{F}_{\theta}\rangle_{\Psi}\langle I\otimes
\hat{p}_{j}\rangle_{\Psi}{}\nonumber\\
&=& N^{2}[\frac{\theta}{4}p_{0j}\epsilon_{ik}p_{0k}-b_{i}p_{0j} - 16
\exp(\frac{-(p_{01}^2 + p_{02}^2)}{2\alpha})
\exp(\frac{-8\alpha(b_{1}^2 + b_{2}^2)}{(\alpha^2\theta^2 +
  4)})\times{}\nonumber\\
&& [\{-\theta^{3}\alpha^{4}b_{i} \epsilon_{jk} b_{k} +
  \theta^{2}\alpha^{2}p_{0i} b_{j} + 4\theta\alpha^{2}
  b_{j}\epsilon_{ik} b_{k} + 4p_{0i}b_{j}\}\{(\alpha^2 \theta^2 + 4)^{-3}\}]-
{}\nonumber\\
&&N^{2}(\frac{-\theta}{2}\epsilon_{ik}p_{0k} + \frac{8\alpha^{2}
  \theta^{2}b_{i}} {(\alpha^{2} \theta^{2}+4)^{2}}
\exp(\frac{-(p_{01}^2 + p_{02}^2)}{2\alpha})
\exp(\frac{-8\alpha(b_{1}^2 + b_{2}^2)}{(\alpha^2 \theta^2+4)}))
\times{}\nonumber\\
&+&\frac{16\alpha^{2}\theta \epsilon_{jk}b_{k}}{(\alpha^{2} \theta^{2}
  + 4)^{2}} \exp(\frac{-(p_{01}^2+p_{02}^2)}{2\alpha}) \exp(\frac{-8
  \alpha(b_{1}^2 + b_{2}^2)}{(\alpha^2 \theta^2 + 4)}))], \\
\bar{G}_{ij} &=& \langle \hat{F}_{\theta}^{-1}(\hat{p}_{i}\otimes \hat{p}_{j})\hat{F}_{\theta}
\rangle_{\Psi}-\langle \hat{p}_{i}\otimes I \rangle_{\Psi}\langle I\otimes
\hat{p}_{j}\rangle_{\Psi}{}\nonumber\\
&=& N^{2}[\frac{-p_{0i}p_{0j}}{2} + 32\exp(\frac{-(p_{01}^2 +
    p_{02}^2)}{2\alpha})exp(\frac{-8\alpha(b_{1}^2 +
    b_{2}^2)}{(\alpha^2 \theta^2 + 4)}) [\{\theta^{2}\alpha^{4}
    \epsilon_{ik} b_{k}\epsilon_{jl}b_{l} + 4\alpha^{2}b_{i}b_{j}\}
{}\nonumber\\
&&\{(\alpha^2\theta^2 + 4)^{-3}\}]-N^{2}(+\frac{16\alpha^{2} \theta
\epsilon_{ik}b_{k}}{(\alpha^{2}\theta^{2} + 4)^{2}}
  \exp(\frac{-(p_{01}^2 + p_{02}^2)}{2\alpha})
  \exp(\frac{-8\alpha(b_{1}^2 + b_{2}^2)}{(\alpha^2 \theta^2 +
    4)})){}\nonumber\\
&&(\frac{16\alpha^{2}\theta \epsilon_{jk} b_{k}}{(\alpha^{2}
    \theta^{2}+4)^{2}} \exp(\frac{-(p_{01}^2 + p_{02}^2)}{2\alpha})]
\exp(\frac{-8\alpha(b_{1}^2+b_{2}^2)}{(\alpha^2\theta^2+4)})), \\
\bar{E}_{ij} &=& \langle \hat{p}_{i}\hat{p}_{j}\otimes I\rangle_{\Psi}-\langle
\hat{p}_{i}\otimes I\rangle_{\Psi}\langle \hat{p}_{j}\otimes
I\rangle_{\Psi}=\langle I\otimes \hat{p}_{i}\hat{p}_{j}\rangle_{\Psi}-\langle
I\otimes \hat{p}_{i})\rangle_{\Psi}\langle I\otimes
\hat{p}_{j}\rangle_{\Psi}{}\nonumber\\
&=& N^{2}[\frac{p_{0i}p_{0j}}{2}+\alpha\delta_{ij}+ 16
  \exp(\frac{-(p_{01}^2 + p_{02}^2)}{2\alpha})
  \exp(\frac{-8\alpha(b_{1}^2 + b_{2}^2)}{(\alpha^2\theta^2+4)})
  \times{}\nonumber\\
&&[\{\theta^{2}\alpha^{3}(2\alpha\epsilon_{ik}b_{k} \epsilon_{jl}
    b_{l} + \delta_{ij}) - 4\alpha(2\alpha b_{i}b_{j} -
    \delta_{ij})\}\{(\alpha^2\theta^2 + 4)^{-3}\}]-{}\nonumber\\
&&N^{2}(+\frac{16\alpha^{2}\theta \epsilon_{ik}b_{k}}
{(\alpha^{2}\theta^{2}+4)^{2}}\exp(\frac{-(p_{01}^2 + p_{02}^2)}{2\alpha})
\exp(\frac{-8\alpha(b_{1}^2 + b_{2}^2)}{(\alpha^2\theta^2 +
  4)}))\times{}\nonumber\\
&&(\frac{16\alpha^{2}\theta \epsilon_{jk}b_{k}}{(\alpha^{2}
  \theta^{2}+4)^{2}} \exp(\frac{-(p_{01}^2 + p_{02}^2)}{2\alpha})
\exp(\frac{-8\alpha(b_{1}^2 + b_{2}^2)}{(\alpha^2 \theta^2+4)}))]
\end{eqnarray}
where the normalization constant is given by
\begin{eqnarray}
N^{-2} = 2(1+\frac{4}{(\alpha^{2}\theta^{2}+4)}
\exp(\frac{-(p_{01}^2+p_{02}^2)}{2\alpha}) \exp(\frac{-8\alpha(b_{1}^2
  + b_{2}^2)}{(\alpha^2\theta^2+4)}))
\end{eqnarray}

\subsection*{Appendix III: Method of the calculation of various integrals}

The method followed to calculate the terms like
$Re{\langle((\hat{\overline{x_{1}}})^{2}\otimes
I)\hat{F}_{\theta}^{-2}\hat{\tau}_{0}\rangle_{\psi_{1}\otimes\psi_{2}}}$ required
to obtain the expectation values is as follows.  The generic element
looks like $\langle(\hat{A}\otimes \hat{B}) \hat{F}_{\theta}^{-2}
\tau_{0}\rangle_{\psi_{1}\otimes\psi_{2}}$. The above expectation
value can be calculated by inserting the complete set of momentum
eigenstates. Therefore,
\begin{eqnarray}
&&\langle(\hat{A}\otimes
\hat{B})\hat{F}_{\theta}^{-2}\hat{\tau}_{0}\rangle_{\psi_{1}\otimes\psi_{2}}=\int
d^{2}kd^{2}l~\exp(i\theta_{ij}k_{i}l_{j})~ \langle\vec{k}|\hat{A}|\psi_{1}
\rangle^{*}\psi_{2}(\vec{k})
\langle\vec{l}|\hat{B}|\psi_{2}\rangle^{*}\psi_{1}(\vec{l}){}\nonumber\\
&=&\int d^{2}kd^{2}l~\exp(ik_{i}q_{i})~ f(\vec{k})g(\vec{l})
\label{ApII1}
\end{eqnarray}
where we define $\theta\epsilon_{ij}l_{j}=q_{i}$,
~$f(\vec{k})=\langle\vec{k}|\hat{A}|\psi_{1}\rangle^{*}\psi_{2}(\vec{k})$,
$g(\vec{l})=\langle\vec{l}|\hat{B}|\psi_{2}\rangle^{*}\psi_{1}(\vec{l})$.
Further, Eq.(\ref{ApII1}) can be simplified as
\begin{eqnarray}
\int
d^{2}kd^{2}l~ \exp(ik_{i}q_{i})~f(\vec{k})g(\vec{l}) =
\frac{2\pi}{\theta^{2}}\int d^{2}k
f(\vec{k})\widetilde{\widetilde{g}}(\vec{k})
\end{eqnarray}
where $d^{2}l=\frac{d^{2}q}{\theta^{2}}$ and
$\widetilde{\widetilde{g}}(\vec{k})$ is the Fourier transform of
$\widetilde{g}(\vec{q})=g(\vec{l})$ given by
\begin{eqnarray}
\widetilde{\widetilde{g}}(\vec{k})=\frac{1}{2\pi}\int
d^{2}q~\exp(ik_{i}q_{i})~\widetilde{g}(\vec{q})
\end{eqnarray}


\begin{thebibliography}{99}
\bibitem{footnote1} Note that we are using a `bar' over $x_{i}$ as
$\bar{x}_{i}$ to indicate a non-commutative variable in contrast
to commutative variables. A further `hat' is added to indicate the
corresponding operator as $\hat{\bar{x}}_{i}$.
\bibitem{footnote4}
Note that $\overline{S}$ is also an element of $Sp(4,\Re)$, as it provides only
a non-standard
representation of $Sp(4,\Re)$, as $\overline{S}$ and $S$
are related by a similarity transformation, somewhat akin to the representation
of the rotation group in a plane using a non-orthonormal basis.
\end{thebibliography}
\end{document}